\documentclass{eas}
\usepackage{graphicx}
\usepackage{bm}  
\usepackage{amssymb}
\newcommand{\R}{\mathbb{R}}
\newcommand{\M}{\mathcal{M}}
\newcommand{\T}{\mathcal{T}}
\newcommand{\w}[1]{\bm{#1}}
\newcommand{\vv}[1]{\vec{\w{#1}}}
\newcommand{\uu}[1]{\underline{\w{#1}}}
\newcommand{\dd}{\w{\mathrm{d}}}
\newcommand{\be}{\begin{equation}}
\newcommand{\ee}{\end{equation}}
\newcommand{\bea}{\begin{eqnarray}}
\newcommand{\eea}{\end{eqnarray}}
\newcommand{\Lie}[1]{\bm{\mathcal L}_{\vv{#1}}\,}
\newcommand{\Liec}[1]{{\mathcal L}_{\vv{#1}}\,}
\newcommand{\der}[2]{\frac{\partial #1}{\partial #2}}
\newcommand{\encadre}[1]{\fbox{$\displaystyle #1$}}

\begin{document}

\title{An introduction to relativistic hydrodynamics}\thanks{to appear in the proceedings
of the School \emph{Astrophysical Fluid Dynamics} (Carg\`ese, France, 9-13 May 2005) organized by B. Dubrulle \& M. Rieutord (EDP Sciences, in press)} 
\author{Eric Gourgoulhon}
\address{Laboratoire de l'Univers et de ses Th\'eories (LUTH),
UMR 8102 du CNRS, Observatoire de Paris, F-92195 Meudon Cedex, France;
\email{eric.gourgoulhon@obspm.fr}}
\begin{abstract}
This lecture provides some introduction to perfect fluid dynamics within the
framework of general relativity. The presentation is based on the
Carter-Lichnerowicz approach. 
It has the advantage over the more traditional
approach of leading very straightforwardly to important conservation laws, such as the relativistic generalizations of
Bernoulli's theorem or Kelvin's circulation theorem. 
It also permits to get easily first integrals of motion which are particularly useful
for computing equilibrium configurations of relativistic stars 
in rotation or in binary systems.
The presentation is relatively self-contained and does not require any a priori
knowledge of general relativity. In particular, the three types of 
derivatives involved in relativistic hydrodynamics are introduced
in detail: this concerns the Lie, exterior and covariant derivatives.
\end{abstract}
\maketitle

\section{Introduction}

Relativistic fluid dynamics is an important topic of modern astrophysics
in at least three contexts: 
(i) jets emerging at relativistic speed from the core of active galactic 
nuclei or from microquasars, and certainly from gamma-ray burst central
engines, (ii) compact stars and flows around black holes,
and (iii) cosmology.
Notice that for items (ii) and (iii) general relativity is necessary,
whereas special relativity is sufficient for (i).

We provide here an introduction to relativistic perfect fluid dynamics in the framework of general relativity, so that it is applicable to all themes (i) to (iii). 
However, we shall make a limited usage of general relativistic concepts. In particular,
we shall not use the Riemannian curvature and all the results will 
be independent of the Einstein equation. 

We have chosen to introduce relativistic hydrodynamics via an 
approach developed originally by Lichnerowicz (\cite{Lichn41}, \cite{Lichn55}, \cite{Lichn67}) and
extended significantly by Carter (\cite{Carte73}, \cite{Carte79}, \cite{Carte89}).
This formulation is very elegant and permits an easy derivation of 
the relativistic generalizations of all the 
standard conservation laws of classical fluid mechanics.
Despite of this, it is absent from most (all ?) textbooks.
The reason may be that the mathematical settings of Carter-Lichnerowicz approach
is Cartan's exterior calculus, which departs from what physicists call ``standard
tensor calculus''. Yet Cartan's exterior calculus is
simpler than the ``standard tensor calculus'' for it does not 
require any specific structure on the spacetime manifold. In particular, it is 
independent of the metric tensor and its associated covariant derivation, not 
speaking about the Riemann curvature tensor. Moreover it is well adapted to the
computation of integrals and their derivatives, a feature which is obviously 
important for hydrodynamics.

Here we start by introducing the
general relativistic spacetime as a pretty simple mathematical structure
(called \emph{manifold}) on which one can define
vectors and multilinear forms. The latter ones map vectors to real numbers, 
in a linear way.
The differential forms on which Cartan's exterior calculus is based are 
then simply multilinear forms that are fully antisymmetric.
We shall describe this in Sec.~\ref{s:fields_deriv}, where we put a special emphasis
on the definition of the three kinds of derivative useful for hydrodynamics:
the \emph{exterior derivative} which acts only on differential forms, the 
\emph{Lie derivative}
along a given vector field and the \emph{covariant derivative} which is associated with
the metric tensor. Then in Sec.~\ref{s:worldlines} we move to physics by
introducing the notions of particle worldline, proper time, 4-velocity and
4-acceleration, as well as Lorentz factor between two observers.
The hydrodynamics then starts in Sec.~\ref{s:fluid_stress_energy} where we introduce
the basic object for the description of a fluid: a bilinear form called the
\emph{stress-energy tensor}. In this section, we define also the concept of 
\emph{perfect fluid}
and that of \emph{equation of state}. The equations of fluid motion are then deduced from 
the local conservation of energy and momentum in Sec.~\ref{s:conser_ener_mom}.
They are given there in the standard form which is essentially a relativistic version
of Euler equation. From this standard form, we derive the 
Carter-Lichnerowicz equation of motion in Sec.~\ref{s:CL_eom}, 
before specializing it 
to the case of an equation of state which depends on two parameters: the baryon number density and the entropy density. 
We also show that the Newtonian limit of the 
Carter-Lichnerowicz equation is a well known alternative form of the Euler equation,
namely the Crocco equation. The power of the Carter-Lichnerowicz approach appears in 
Sec.~\ref{s:conserv_theor} where we realize how easy it is to derive 
conservation laws from it, among which the relativistic version of the classical 
Bernoulli theorem and Kelvin's circulation theorem. We also show that some of these
conservation laws are useful for getting numerical solutions for rotating relativistic 
stars or relativistic binary systems.


\section{Fields and derivatives in spacetime} \label{s:fields_deriv}

It is not the aim of this lecture to provide an introduction to general
relativity. For this purpose we refer the reader to two excellent introductory textbooks which have 
recently appeared: (Hartle \cite{Hartl03}) and (Carrol \cite{Carro04}).
Here we recall only some basic geometrical concepts which are fundamental 
to a good understanding of relativistic hydrodynamics.
In particular we focus on the various notions of derivative on spacetime.

\subsection{The spacetime of general relativity} \label{s:curved_spacetime}

Relativity has performed the fusion of \emph{space} and \emph{time}, two notions
which were completely distinct in Newtonian mechanics. This gave rise to 
the concept of \emph{spacetime}, on which both the special and general theory 
of relativity are based. Although this is not particularly fruitful (except for
contrasting with the relativistic case), one may also speak of spacetime 
in the Newtonian framework. The Newtonian spacetime $\M$ is then nothing but the 
affine space $\R^4$, foliated by the hyperplanes $\Sigma_t$ 
of constant absolute time $t$:
these hyperplanes represent the ordinary 3-dimensional space at successive instants.
The foliation $(\Sigma_t)_{t\in\R}$
is a basic structure of the Newtonian spacetime and does not depend
upon any observer. The \emph{worldline} $\mathcal{L}$ of a particle is the curve 
in $\M$ generated by the successive positions of the particle.
At any point $A\in\mathcal{L}$, 
the time read on a clock moving along $\mathcal{L}$ is simply the
parameter $t$ of the hyperplane $\Sigma_t$ that intersects $\mathcal{L}$
at $A$.

The spacetime $\mathcal{M}$ of special relativity is the 
same mathematical space as the
Newtonian one, i.e. the affine space $\R^4$. The major difference with
the Newtonian case is that there does not exist any privileged foliation
$(\Sigma_t)_{t\in\R}$. Physically this means that the notion of
absolute time is absent in special relativity. 
However $\M$ is still endowed with some absolute structure: 
the \emph{metric tensor} $\w{g}$ and the
associated \emph{light cones}. The metric tensor is a symmetric bilinear form $\w{g}$
on $\M$, which defines the scalar product of vectors. The null (isotropic) 
directions of $\w{g}$ give the worldlines of photons (the light cones). Therefore
these worldlines depend only on the absolute structure $\w{g}$ and not, for instance,
on the observer who emits the photon.

The spacetime $\mathcal{M}$ of general relativity differs from both Newtonian and special relativistic spacetimes, in so far as it is no longer the affine space
$\R^4$ but a more general mathematical structure, namely a \emph{manifold}.
A \emph{manifold of dimension 4} is a topological space such that around each point 
there exists a neighbourhood which is 
homeomorphic to an open subset of $\R^4$.
This simply means that, locally, one can label the points of $\M$ in a
continuous way by 4 real numbers $(x^\alpha)_{\alpha\in\{0,1,2,3\}}$
(which are called \emph{coordinates}). To cover the full $\M$, several different
coordinates patches (\emph{charts} in mathematical jargon) can be required. 

\begin{figure}
\centerline{\includegraphics[width=0.7\textwidth]{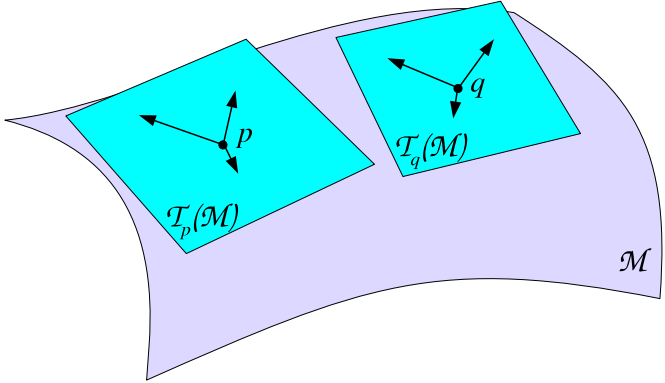}}
\caption{\label{f:esp_tangent} The vectors at two points $p$ and $q$ on the
spacetime manifold $\M$ belong to two different vector spaces: 
the tangent spaces $\T_p(\M)$ and $\T_q(\M)$.}
\end{figure}

Within the manifold structure the definition of vectors is not as trivial
as within the affine structure of the Newtonian and special relativistic
spacetimes. Indeed, only infinitesimal vectors connecting 
two infinitely close points can be defined a priori on a manifold. At a given
point $p\in\M$, the set of such vectors generates a 4-dimensional vector
space, which is called the \emph{tangent space} at the point $p$ and is
denoted by $\T_p(\M)$. The situation is therefore different from the Newtonian
or special relativistic one, for which the very definition of an affine space
provides a unique global vector space. On a manifold
there are as many vector spaces as points $p$ (cf. Fig.~\ref{f:esp_tangent}).

Given a vector basis $(\vv{e}_\alpha)_{\alpha\in\{0,1,2,3\}}$ of $\T_p(\M)$, the components of a vector $\vv{v}\in\T_p(\M)$ on this basis are denoted by $v^\alpha$:
$\vv{v} = v^\alpha \vv{e}_\alpha$, where we have employed Einstein's convention 
for summation on repeated indices. 
It happens frequently that the vector basis is associated to a coordinate
system $(x^\alpha)$ on the manifold, in the following way. If $dx^\alpha$ is the
(infinitesimal) difference of coordinates between $p$ and a neighbouring point
$q$, the components of the vector $\overrightarrow{pq}$ with respect to 
the basis $(\vv{e}_\alpha)$ are exactly $dx^\alpha$. The basis that fulfills this
property is unique and is called the \emph{natural basis} associated with the coordinate
system $(x^\alpha)$. It is usually denoted by $(\partial/\partial x^\alpha)$,
which is a reminiscence of the intrinsic definition of vectors on a manifold
as differential operators acting on scalar fields.

As for special relativity, the absolute structure given on the spacetime manifold $\M$ 
of general relativity is the \emph{metric tensor} $\w{g}$. 
It is now a field on $\M$: at each point $p\in\M$, $\w{g}(p)$ is a symmetric
bilinear form acting on vectors in the tangent space $\T_p(\M)$:
\be 
	\begin{array}{rccl}
	\w{g}(p): & \T_p(\M)\times\T_p(\M) & \longrightarrow & \R \\
		& (\vv{u},\vv{v}) & \longmapsto & \w{g}(\vv{u},\vv{v}) 
	=: \vv{u}\cdot\vv{v} .
	\end{array}
\ee
It is demanded that the bilinear form $\w{g}$ is not degenerate and is of signature
$(-,+,+,+)$. It thus defines a \emph{scalar product} on $\T_p(\M)$, which justifies
the notation $\vv{u}\cdot\vv{v}$ for $\w{g}(\vv{u},\vv{v})$.
The isotropic directions of $\w{g}$ give the local \emph{light cones}:
a vector $\vv{v}\in\T_p(\M)$ is tangent to a light cone and called a
\emph{null} or \emph{lightlike} vector iff
$\vv{v}\cdot\vv{v} = 0$. Otherwise, the vector is said to be \emph{timelike}
iff $\vv{v}\cdot\vv{v} < 0$ and \emph{spacelike} iff $\vv{v}\cdot\vv{v} > 0$.

\subsection{Tensors}  \label{s:tensors}

Let us recall that a \emph{linear form} at a given point $p\in\M$ is
an application
\be
	\begin{array}{rccl}
	\w{\omega}: & \T_p(\M) & \longrightarrow & \R \\
		& \vv{v} & \longmapsto & \langle\w{\omega},\vv{v}\rangle :=
			\w{\omega}(\vv{v})  ,
	\end{array}
\ee
that is linear. The set of all linear forms at $p$ forms a vector space of dimension
4, which is denoted by $\T_p(\M)^*$ and is called the \emph{dual} of the 
tangent space $\T_p(\M)$. In relativistic physics, an abundant use is made
of linear forms and their generalizations: the tensors.
A \emph{tensor of type} $(k,\ell)$, also called \emph{tensor $k$ times
contravariant and $\ell$ times covariant}, is an application 
\be \label{e:def_tensor}
	\begin{array}{rccl}
	\w{T}: & \underbrace{\T_p(\M)^*\times\cdots\times\T_p(\M)^*}_{k {\ \rm times}}
	\times \underbrace{\T_p(\M)\times\cdots\times\T_p(\M)}_{\ell {\ \rm times}}
	& \longrightarrow & \R  \\
	& (\w{\omega}_1,\ldots,\w{\omega}_k,\vv{v}_1,\ldots,\vv{v}_\ell) 
		& \longmapsto & 
	\w{T}(\w{\omega}_1,\ldots,\w{\omega}_k,\\
	& & &  \ \quad  \vv{v}_1,\ldots,\vv{v}_\ell)
	\end{array}
\ee
that is linear with respect to each of its arguments. The integer $k+\ell$ is
called the \emph{valence} of the tensor. Let us recall the canonical duality
$\T_p(\M)^{**}=\T_p(\M)$, which means that every vector $\vv{v}$ can be considered
as a linear form on the space $\T_p(\M)^*$, defining the application
$\vv{v}:\; \T_p(\M)^*\rightarrow \R$, 
$\w{\omega}\mapsto \langle\w{\omega},\vv{v}\rangle$.
Accordingly a vector is a tensor of type $(1,0)$. A linear form is a
tensor of type $(0,1)$ and the metric tensor $\w{g}$
is a tensor of type $(0,2)$.

Let us consider a vector
basis of $\T_p(\M)$, $(\vv{e}_\alpha)$, which can be either a natural
basis (i.e. related to some coordinate system) or not (this is often the case for
bases orthonormal with respect to the metric $\w{g}$). There exists then a unique
quadruplet of 1-forms, $(\w{e}^\alpha)$, that constitutes a basis of the dual space
$\T_p(\M)^*$ and that verifies
\be
	\langle\w{e}^\alpha,\vv{e}_\beta\rangle = \delta^\alpha_{\ \, \beta} ,
\ee
where $\delta^\alpha_{\ \, \beta}$ is the Kronecker symbol.
Then we can expand any tensor $\w{T}$ of type $(k,\ell)$ as
\be \label{e:def_components}
	\encadre{\w{T} = T^{\alpha_1\ldots\alpha_k}_{\qquad\ \; \beta_1\ldots\beta_\ell}
		\; \vv{e}_{\alpha_1} \otimes \ldots \otimes \vv{e}_{\alpha_k} 
                \otimes
		\w{e}^{\beta_1} \otimes \ldots \otimes \w{e}^{\beta_\ell} } ,
\ee
where the \emph{tensor product} $ \vv{e}_{\alpha_1} \otimes \ldots \otimes \vv{e}_{\alpha_k} \otimes
\w{e}^{\beta_1} \otimes \ldots \otimes \w{e}^{\beta_\ell}$ is the tensor of
type $(k,\ell)$ for which the image of  $(\w{\omega}_1,\ldots,\w{\omega}_k,\vv{v}_1,\ldots,\vv{v}_\ell)$ as in 
(\ref{e:def_tensor}) is the real number
\be \label{e:def_tensor_product}
	\prod_{i=1}^k \langle\w{\omega}_i,\vv{e}_{\alpha_i}\rangle \;\times\; 
	\prod_{j=1}^\ell \langle\w{e}^{\beta_j},\vv{v}_j\rangle .
\ee
Notice that all the products in the above formula are simply products in $\R$.
The $4^{k+\ell}$ scalar coefficients  $T^{\alpha_1\ldots\alpha_k}_{\qquad\ \; \beta_1\ldots\beta_\ell}$ in (\ref{e:def_components}) are called the \emph{components
of the tensor $\w{T}$ with respect to the basis} $(\vv{e}_\alpha)$, or \emph{with
respect to the coordinates $(x^\alpha)$} if $(\vv{e}_\alpha)$ is the natural basis
associated with these coordinates. These components are unique and fully characterize
the tensor $\w{T}$. Actually, in many studies, a basis is assumed (mostly a natural 
basis) and the tensors are always represented by their components. 
This way of presenting things is called the \emph{index notation}, or the \emph{abstract index notation} if the basis is not specified (e.g. Wald \cite{Wald84}). We shall not
use it here, sticking to what is called the \emph{index-free notation} and which is
much better adapted to exterior calculus and Lie derivatives.

The notation $v^\alpha$ already introduced for the components of a vector $\vv{v}$ is of course the particular case $(k=1,\ell=0)$ of the general definition given above.
For a linear form $\w{\omega}$, the components $\omega_\alpha$ are such that 
$\w{\omega}=\omega_\alpha \w{e}^\alpha$ [Eq.~(\ref{e:def_components}) with
$(k=0,\ell=1)$]. Then
\be
	\langle\w{\omega},\vv{v}\rangle = \omega_\alpha v^\alpha . 
\ee
Similarly the components $g_{\alpha\beta}$ of the metric tensor $\w{g}$ are defined
by $\w{g}=g_{\alpha\beta} \, \w{e}^\alpha\otimes \w{e}^\beta$ 
[Eq.~(\ref{e:def_components}) with $(k=0,\ell=2)$] and the scalar products are
expressed in terms of the components as
\be
	\w{g}(\vv{u},\vv{v}) = g_{\alpha\beta} u^\alpha v^\beta .
\ee

\subsection{Scalar fields and their gradients} \label{s:scalar_gradient}

A \emph{scalar field} on the spacetime manifold $\M$ is an application
$f:\  \M\rightarrow \R$. If $f$ is smooth, it gives rise
to a field of linear forms (such fields are called  \emph{1-forms}), 
called the \emph{gradient of} $f$ and denoted $\dd f$.
It is defined so that the variation of $f$ between two neighbouring points $p$ and $q$
is\footnote{do not confuse the increment $df$ of $f$ with the gradient 1-form
$\dd f$: the boldface \emph{d} is used to distinguish the latter from the former}
\be \label{e:df}
	df = f(q) - f(p) =  \langle \dd f, \overrightarrow{pq} \rangle .
\ee 
Let us note that, in non-relativistic physics, the gradient is very often 
considered as a vector and not as a 1-form. 
This is because one associates implicitly a vector
$\vec{\omega}$ to any 1-form $\omega$ thanks to the Euclidean scalar product 
of $\R^3$,
via $\forall \vec{v}\in \R^3,\ \langle\omega,\vec{v}\rangle = \vec{\omega}\cdot\vec{v}$.
Accordingly, the formula (\ref{e:df}) is rewritten as
$df = \vec{\nabla} f \cdot \overrightarrow{pq}$. But one shall keep in mind
that, fundamentally, the gradient is a 1-form and not a vector.

If $(x^\alpha)$ is a coordinate system on $\M$ and 
$(\vv{e}_\alpha = \partial/\partial x^\alpha)$ the associated natural basis,
then the dual basis is constituted by the gradients of the four coordinates:
$\w{e}^\alpha = \dd x^\alpha$. The components of the gradient of any scalar field $f$ in
this basis are then nothing but the partial derivatives of $f$:
\be
	\dd f = (\dd f)_\alpha \,  \dd x^\alpha
	\qquad \mbox{with} \qquad (\dd f)_\alpha = \der{f}{x^\alpha} .
\ee

\subsection{Comparing vectors and tensors at different spacetime points: various derivatives on $\M$}

A basic concept for hydrodynamics is of course that of \emph{vector field}.
On the manifold $\M$, this means the choice of a 
vector $\vv{v}(p)$ in $\T_p(\M)$ for each $p\in\M$.
We denote by $\T(\M)$ the space of all smooth vector fields on $\M$
\footnote{The experienced reader is warned that $\T(\M)$ does not stand 
for the tangent bundle of $\M$ (it rather corresponds to the 
space of smooth cross-sections of that bundle). No confusion may arise since 
we shall not use the notion of bundle.}.
The derivative of the vector field is to be constructed for the variation $\delta\vv{v}$
of $\vv{v}$ between two neighbouring points $p$ and $q$. 
Naively, one would write $\delta\vv{v} = \vv{v}(q)-\vv{v}(p)$, as in
(\ref{e:df}).
However $\vv{v}(q)$ and $\vv{v}(p)$ belong to different vector spaces:
$\T_q(\M)$ and $\T_p(\M)$ (cf. Fig.~\ref{f:esp_tangent}).
Consequently the subtraction $\vv{v}(q)-\vv{v}(p)$ is ill defined,
contrary of the subtraction of two real numbers in (\ref{e:df}).
To proceed in the definition of the derivative of a vector field, one must
introduce some extra-structure on the manifold $\M$: this can be either
another vector field $\vv{u}$, leading to the derivative of $\vv{v}$ along $\vv{u}$
which is called the \emph{Lie derivative}, or a \emph{connection} $\w{\nabla}$ (usually
related to the metric tensor $\w{g}$), leading to the \emph{covariant derivative}
$\w{\nabla}\vv{v}$. These two types of derivative generalize straightforwardly to
any kind of tensor field. For the specific kind of tensor fields constituted by
differential forms, there exists a third type of derivative, which does not 
require any extra structure on $\M$: the \emph{exterior derivative}. 
We will discuss the latter in Sec.~\ref{s:diff_forms}. In the current section, 
we shall review successively the Lie and covariant derivatives.

\begin{figure}
\centerline{\includegraphics[width=0.6\textwidth]{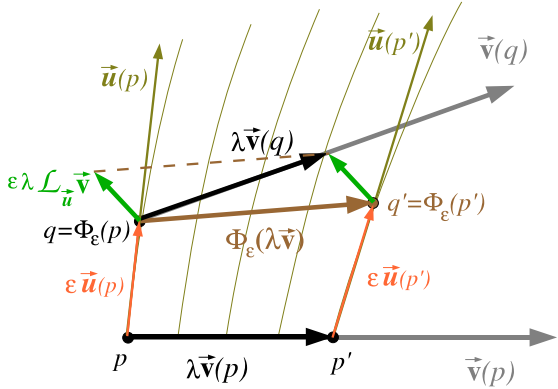}}
\caption{\label{f:lie_deriv} Geometrical construction of the Lie derivative of a
vector field: given a small parameter $\lambda$, each extremity of the arrow
$\lambda\vv{v}$ is dragged by some small parameter $\varepsilon$ 
along $\vv{u}$, to form
the vector denoted by $\Phi_\varepsilon(\lambda\vv{v})$. The latter is then compared with
the actual value of $\lambda\vv{v}$ at the point $q$, the difference (divided 
by $\lambda\varepsilon$) defining the Lie derivative $\Lie{u}\vv{v}$.}
\end{figure}

\subsubsection{Lie derivative} \label{s:lie_deriv}

The Lie derivative is a very natural operation in the context of fluid mechanics.
Indeed, consider a vector field $\vv{u}$ on $\M$, called hereafter the \emph{flow}.
Let $\vv{v}$ be another vector field on $\M$, the variation of which is to be studied.
We can use the flow $\vv{u}$ to transport the vector $\vv{v}$ from one point $p$ to
a neighbouring one $q$ and then define rigorously the variation of $\vv{v}$
as the difference between the actual value of $\vv{v}$ at $q$ and the transported
value via $\vv{u}$. More precisely the definition of the Lie derivative of 
$\vv{v}$ with respect to $\vv{u}$ is as follows (see Fig.~\ref{f:lie_deriv}).
We first define the image $\Phi_\varepsilon(p)$ of the point $p$ by the transport by an infinitesimal ``distance'' $\varepsilon$ along the field lines of $\vv{u}$ as 
$\Phi_\varepsilon(p)=q$, where $q$ is the point close to $p$ such that
$\overrightarrow{pq}=\varepsilon\vv{u}(p)$.
Besides, if we multiply the vector $\vv{v}(p)$ by 
some infinitesimal parameter $\lambda$, it becomes an infinitesimal vector at $p$.
Then there exists a unique point $p'$ close to $p$ such that 
$\lambda\vv{v}(p)=\overrightarrow{pp'}$.
We may transport the point $p'$ to a point $q'$ along the field lines of
$\vv{u}$ by the same ``distance'' $\varepsilon$ as that used to transport
$p$ to $q$: $q'=\Phi_\varepsilon(p')$ (see Fig.~\ref{f:lie_deriv}). $\overrightarrow{qq'}$ is then an
infinitesimal vector at $q$ and we
define the transport by the distance $\varepsilon$ of the vector $\vv{v}(p)$ 
along the field lines of $\vv{u}$ according to
\be
	\Phi_\varepsilon(\vv{v}(p)) := \frac{1}{\lambda} \, \overrightarrow{qq'}.
\ee
$\Phi_\varepsilon(\vv{v}(p))$ is vector in $\T_q(\M)$. We may then subtract it from the
actual value of the field $\vv{v}$ at $q$ and define the \emph{Lie derivative}
of $\vv{v}$ along $\vv{u}$ by
\be
	\Lie{u} \vv{v} := \lim_{\varepsilon\rightarrow 0} \frac{1}{\varepsilon}
	\left[ \vv{v}(q) - \Phi_\varepsilon(\vv{v}(p)) \right] .
\ee

If we consider a coordinate system $(x^\alpha)$ adapted to the
field $\vv{u}$ in the sense that $\vv{u}=\vv{e}_0$ where $\vv{e}_0$ is the first
vector of the natural basis associated with the coordinates $(x^\alpha)$, then
the Lie derivative is simply given by the partial derivative of the vector components
with respect to $x^0$:
\be \label{e:Lie_adapted}
	\left( \Lie{u} \vv{v} \right)^\alpha = \der{v^\alpha}{x^0} .
\ee
In an arbitrary coordinate system, this formula is generalized to 
\be
	\Liec{u} v^\alpha = u^\mu \der{v^\alpha}{x^\mu}
	- v^\mu \der{u^\alpha}{x^\mu} , 
\ee
where use has been made of the standard notation 
$\Liec{u} v^\alpha := \left( \Lie{u} \vv{v} \right)^\alpha$.

The Lie derivative is extended to any tensor field by (i) demanding that for
a scalar field $f$, $\Lie{u} f = \langle\dd f,\vv{u}\rangle$ and (ii) using the Leibniz
rule. As a result, the Lie derivative $\Lie{u}\w{T}$ of a tensor field $\w{T}$ of type 
$(k,\ell)$ is a tensor field of the same type, the components of which
with respect to a given coordinate system $(x^\alpha)$ are
\bea
\Liec{u} T^{\alpha_1\ldots\alpha_k}_{\qquad\ \; \beta_1\ldots\beta_\ell}&=&
u^\mu \der{}{x^\mu} T^{\alpha_1\ldots\alpha_k}_{\qquad\ \; \beta_1\ldots\beta_\ell} 
- \sum_{i=1}^k T^{\alpha_1\ldots
\!{{{\scriptstyle i\atop\downarrow}\atop \scriptstyle\sigma}\atop\ }\!\!
\ldots\alpha_k}_{\qquad\ \ \ \  \  \  \beta_1\ldots\beta_\ell}
 \; \der{u^{\alpha_i}}{x^\sigma} \nonumber \\
& & +  \sum_{i=1}^\ell T^{\alpha_1\ldots\alpha_k}_{\qquad\ \; \beta_1\ldots
\!{\ \atop {\scriptstyle\sigma \atop {\uparrow\atop \scriptstyle i}} }\!\!
\ldots\beta_\ell} 
\; \der{u^{\sigma}}{x^{\beta_i}} . \label{e:Lie_der_comp}
\eea 
In particular, for a 1-form,
\be \label{e:Lie_der_1form}
	\Liec{u} \omega_\alpha = u^\mu \der{\omega_\alpha}{x^\mu}
	+ \omega_\mu \der{u^\mu}{x^\alpha} .
\ee

\subsubsection{Covariant derivative} \label{s:cov_deriv}

The variation $\delta\vv{v}$ of the vector field $\vv{v}$ between two neighbouring
points $p$ and $q$ can be defined if some \emph{affine connection} is given on the
manifold $\M$. The latter is an operator
\be \label{e:def_aff_connec}
	\begin{array}{cccc}
	\w{\nabla} \ : & \T(\M)\times\T(\M) & \longrightarrow & \T(\M) \\
		& (\vv{u},\vv{v}) & \longmapsto & \w{\nabla}_{\vv{u}} \,\vv{v} 
	\end{array} 
\ee
that satisfies all the properties of a derivative operator (Leibniz rule,
etc...), which we shall not list here (see e.g. Wald \cite{Wald84}). 
The variation of $\vv{v}$ (with respect to
the connection $\w{\nabla}$) between two neighbouring points $p$ and $q$ is then
defined by 
\be
	\delta \vv{v} := \w{\nabla}_{\overrightarrow{pq}} \, \vv{v} .
\ee
One says that $\vv{v}$ is \emph{transported parallelly to itself} between $p$ and $q$
iff $\delta\vv{v}=0$.
From the manifold structure alone, there exists an infinite number of possible 
connections and none is preferred. Taking account the metric tensor $\w{g}$ changes the
situation: there exists a unique connection, called the 
\emph{Levi-Civita connection}, such that the tangent vectors
to the geodesics with respect to $\w{g}$ are transported parallelly to themselves
along the geodesics. In what follows, we will make use only of the Levi-Civita
connection.

Given a vector field $\vv{v}$ and a point $p\in\M$, we can consider the
type $(1,1)$ tensor at $p$ denoted by $\w{\nabla}\vv{v}(p)$ and defined by
\be \label{e:def_nabv_p}
	\begin{array}{cccc}
	\w{\nabla}\vv{v}(p) \ : & \T_p(\M)^*\times\T_p(\M) & \longrightarrow & \R \\
		& (\w{\omega},\vv{u}) & \longmapsto & 
	\left\langle \w{\omega},\,  (\w{\nabla}_{\vv{u}_{\rm c}} \, \vv{v})(p) \right\rangle
	\end{array} , 
\ee
where $\vv{u}_{\rm c}$ is a vector field that performs some extension of the
vector $\vv{u}$ in the neighbourhood of $p$: $\vv{u}_{\rm c}(p)=\vv{u}$.
It can be shown that the map (\ref{e:def_nabv_p})
is independent of the choice of $\vv{u}_{\rm c}$.
Therefore $\w{\nabla}\vv{v}(p)$ is a type $(1,1)$ tensor at $p$ which depends only on the
vector field $\vv{v}$. By varying $p$ we get a type
$(1,1)$ tensor field denoted $\w{\nabla}\vv{v}$ and called the \emph{covariant 
derivative} of $\vv{v}$.

As for the Lie derivative, the covariant derivative is extended to any tensor
field by (i) demanding that for a scalar field
$\w{\nabla} f = \dd f$ and (ii) using the Leibniz rule.
As a result, the covariant derivative of a tensor field $\w{T}$ of type $(k,\ell)$ is
a tensor field $\w{\nabla}\w{T}$ of type $(k,\ell+1)$.
Its components with respect a given coordinate system $(x^\alpha)$
are denoted 
\be
\nabla_\gamma T^{\alpha_1\ldots\alpha_k}_{\qquad\ \; \beta_1\ldots\beta_\ell}
	:= 
(\w{\nabla}\w{T})^{\alpha_1\ldots\alpha_k}_{\qquad\ \; \beta_1\ldots\beta_\ell\gamma}
\ee
(notice the position of the index $\gamma$ !) and are given by
\bea
\nabla_\gamma T^{\alpha_1\ldots\alpha_k}_{\qquad\ \; \beta_1\ldots\beta_\ell}&=&
 \der{}{x^\gamma} T^{\alpha_1\ldots\alpha_k}_{\qquad\ \; \beta_1\ldots\beta_\ell} 
+ \sum_{i=1}^k \Gamma^{\alpha_i}_{\ \, \gamma\sigma}\; T^{\alpha_1\ldots
\!{{{\scriptstyle i\atop\downarrow}\atop \scriptstyle\sigma}\atop\ }\!\!
\ldots\alpha_k}_{\qquad\ \ \ \  \  \  \beta_1\ldots\beta_\ell} \nonumber \\
& & -  \sum_{i=1}^\ell \Gamma^\sigma_{\ \, \gamma\beta_i} \; 
T^{\alpha_1\ldots\alpha_k}_{\qquad\ \; \beta_1\ldots
\!{\ \atop {\scriptstyle\sigma \atop {\uparrow\atop \scriptstyle i}} }\!\!
\ldots\beta_\ell}  ,	\label{e:cov_derivT_comp}
\eea 
where the coefficients $\Gamma^\alpha_{\ \, \gamma\beta}$ are the
\emph{Christoffel symbols} of the metric $\w{g}$ with respect to the coordinates $(x^\alpha)$.
They are expressible in terms of the partial derivatives of the components of 
the metric tensor, via
\be 
	 \Gamma^\alpha_{\ \, \gamma\beta} := \frac{1}{2} g^{\alpha\sigma}
	\left( \der{g_{\sigma\beta}}{x^\gamma} + \der{g_{\gamma\sigma}}{x^\beta}
	- \der{g_{\gamma\beta}}{x^\sigma} \right) . 
\ee

A distinctive feature of the Levi-Civita connection is that
\be \label{e:nabla_g_zero}
	\encadre{ \w{\nabla} \w{g} = 0 }. 
\ee

Given a vector field $\vv{u}$ and a tensor field $\w{T}$ of type $(k,\ell)$, 
we define the \emph{covariant derivative of $\w{T}$ along $\vv{u}$} as 
the generalization of (\ref{e:def_aff_connec}):
\be
	\w{\nabla}_{\vv{u}} \, \w{T} := \w{\nabla}\w{T}
        (\underbrace{.,\ldots,.}_{k+\ell\ {\rm slots}},\vv{u}) .
\ee
Notice that $\w{\nabla}_{\vv{u}} \, \w{T}$ is a tensor of the same type $(k,\ell)$
as $\w{T}$ and that its components are
\be
	\left(\w{\nabla}_{\vv{u}} \, \w{T}
	\right)^{\alpha_1\ldots\alpha_k}_{\qquad\ \; \beta_1\ldots\beta_\ell}
	= u^\mu \nabla_\mu T^{\alpha_1\ldots\alpha_k}_{\qquad\ \; \beta_1\ldots\beta_\ell} .
\ee

\subsection{Differential forms and exterior derivatives} \label{s:diff_forms}

The \emph{differential forms} or \emph{$n$-forms} are type $(0,n)$ 
tensor fields that are antisymmetric in all their arguments.
Otherwise stating, at each point $p\in\M$, they constitute
antisymmetric multilinear forms on the vector space $\T_p(\M)$). 
They play a special role in the theory of integration on a 
manifold. Indeed the primary definition of an integral over a manifold of 
dimension $n$ is the integral of a $n$-form. The 4-dimensional volume element 
associated with the metric $\w{g}$ is a 4-form,
called the Levi-Civita alternating tensor. Regarding physics, it is well known that the
electromagnetic field is fundamentally a 2-form (the Faraday tensor $\w{F}$); besides,
we shall see later that the vorticity of a fluid is described by a 2-form, which 
plays a key role in the Carter-Lichnerowicz formulation.

Being tensor fields, the $n$-forms are subject to the Lie and covariant
derivations discussed above. But, in addition, they are subject to a third type
of derivation, called \emph{exterior derivation}.
The \emph{exterior derivative} of a $n$-form $\w{\omega}$ is a
$(n+1)$-form which is denoted $\dd\w{\omega}$. 
In terms of components with respect to a given
coordinate system $(x^\alpha)$, $\dd\w{\omega}$ is defined by
\bea
	\mbox{0-form (scalar field)} & : & (\dd\w{\omega})_\alpha = 
		\der{\omega}{x^\alpha} \label{e:def_ext_0f} \\
	\mbox{1-form} & : & (\dd\w{\omega})_{\alpha\beta} =
	\der{\omega_\beta}{x^\alpha} - \der{\omega_\alpha}{x^\beta}
			 \label{e:def_ext_1f} \\
	\mbox{2-form} & : & (\dd\w{\omega})_{\alpha\beta\gamma} =
	\der{\omega_{\beta\gamma}}{x^\alpha} + 
	\der{\omega_{\gamma\alpha}}{x^\beta} + 
	\der{\omega_{\alpha\beta}}{x^\gamma} \label{e:def_ext_2f} \\
	\mbox{etc...} 
\eea
It can be easily checked that these formul\ae, although expressed in terms of 
partial derivatives of components in a coordinate system, do define tensor fields.
Moreover, the result is clearly antisymmetric (assuming that $\w{\omega}$ is), so 
that we end up with $(n+1)$-forms.
Notice that for a scalar field (0-form), the exterior derivative is nothing but the
gradient 1-form $\dd f$ already defined in Sec.~\ref{s:scalar_gradient}.
Notice also that the definition of the exterior derivative appeals only to the
manifold structure. It does not depend upon the metric tensor  $\w{g}$, nor upon 
any other extra structure on $\M$. We may also notice that
all partial derivatives in the formul\ae\ 
(\ref{e:def_ext_0f})-(\ref{e:def_ext_2f}) can be replaced by covariant derivatives
(thanks to the symmetry of the Christoffel symbols).

A fundamental property of the exterior derivation is to be nilpotent:
\be \label{e:ext_der_nilpot}
	\encadre{ \dd\dd\w{\omega} = 0 }.
\ee
A $n$-form $\w{\omega}$ is said to be \emph{closed} iff $\dd\w{\omega}=0$,
and \emph{exact} iff there exists a $(n-1)$-form $\w{\sigma}$ such that
$\w{\omega} = \dd\w{\sigma}$. From property (\ref{e:ext_der_nilpot}),
an exact $n$-form is closed. The Poincar\'e lemma states that the converse is true,
at least locally. 

The exterior derivative enters in the well known \emph{Stokes' theorem}: if $\mathcal{D}$
is a submanifold of $\M$ of dimension $d$ ($d\in\{1,2,3,4\}$) that has a boundary (denoted $\partial\mathcal{D}$), then for any $(d-1)$-form $\w{\omega}$,
\be \label{e:Stokes}
	\oint_{\partial\mathcal{D}} \w{\omega} = 
	\int_{\mathcal{D}} \dd\w{\omega} .
\ee
Note that $\partial\mathcal{D}$ is a manifold of dimension $d-1$ and 
$\dd\w{\omega}$ is a $d$-form, so that each side of
(\ref{e:Stokes}) is (of course !) a well defined quantity,
as the integral of a $n$-form over a $n$-dimensional manifold.

Another very important formula where the exterior derivative enters is
the \emph{Cartan identity}, which states that the Lie derivative of a $n$-form 
$\w{\omega}$ along a vector field $\vv{u}$ is expressible as
\be \label{e:Cartan}
	\encadre{ \Lie{u}\w{\omega} = \vv{u}\cdot\dd\w{\omega}
	+ \dd(\vv{u}\cdot\w{\omega}) }.
\ee
In the above formula, a dot denotes the contraction on adjacent indices, i.e.
$\vv{u}\cdot\w{\omega}$ is the $(n-1)$-form $\w{\omega}(\vv{v},.,\ldots,.)$,
with the $n-1$ last slots remaining free. Notice that in 
the case where $\w{\omega}$ is a 1-form, Eq.~(\ref{e:Cartan}) is readily obtained 
by combining Eqs.~(\ref{e:Lie_der_1form}) and (\ref{e:def_ext_1f}). 
In this lecture, we shall make an extensive use of the Cartan identity.


\section{Worldlines in spacetime} \label{s:worldlines}

\subsection{Proper time, 4-velocity and 4-acceleration}

\begin{figure}
\centerline{\includegraphics[height=0.25\textheight]{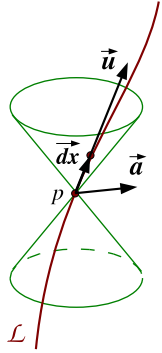}}
\caption{\label{f:worldline} Worldline $\mathcal{L}$ of a particle, with the
4-velocity $\vv{u} = c^{-1} d\vv{x}/d\tau$ and the 4-acceleration $\vv{a}$.}
\end{figure}

A particle or ``point mass'' is fully described by its mass $m>0$ and its worldline 
$\mathcal{L}$ in spacetime. The latter is postulated to be \emph{timelike}, i.e.
such that any tangent vector is timelike. This means that $\mathcal{L}$ lies always
inside the light cone (see Fig.~\ref{f:worldline}).
The \emph{proper time} $d\tau$ corresponding to an elementary displacement\footnote{we denote  by $d\vv{x}$ the infinitesimal vector between two neighbouring points on 
$\mathcal{L}$, but it should be clear that this vector is independent of any
coordinate system $(x^\alpha)$ on $\M$.} $d\vv{x}$
along $\mathcal{L}$ is nothing but the length, as given by the metric tensor, of the
vector $d\vv{x}$ (up to a $c$ factor) :
\be
	c\, d\tau = \sqrt{-\w{g}(d\vv{x},d\vv{x})} .
\ee
The \emph{4-velocity} of the particle is then the vector defined by
\be \label{e:def_4vel}
	\encadre{ \vv{u} := \frac{1}{c} \frac{d\vv{x}}{d\tau} }.
\ee
By construction, $\vv{u}$ is a vector tangent to the worldline $\mathcal{L}$
and is a unit vector with respect to the metric $\w{g}$:
\be
	\vv{u}\cdot\vv{u} = - 1 . 
\ee
Actually, $\vv{u}$ can be characterized as the unique unit tangent vector to
$\mathcal{L}$ oriented toward the future.
Let us stress that the 4-velocity is intrinsic to the particle under consideration:
contrary to the ``ordinary'' velocity, it is not defined relatively to some
observer.

The \emph{4-acceleration} of the particle is the covariant derivative of 
the 4-velocity along itself:
\be \label{e:def_4acc}
	\encadre{ \vv{a} := \w{\nabla}_{\vv{u}} \, \vv{u} }.
\ee
Since $\vv{u}$ is a unit vector, it follows that
\be \label{e:u_ortho_a}
	\vv{u}\cdot\vv{a} = 0 ,
\ee
i.e. $\vv{a}$ is orthogonal to $\vv{u}$ with respect to the metric $\w{g}$
(cf. Fig.~\ref{f:worldline}). In particular, $\vv{a}$ is a spacelike vector.
Again, the 4-acceleration is not relative to any observer, but is intrinsic to
the particle.

\subsection{Observers, Lorentz factors and relative velocities}

Let us consider an observer $\mathcal{O}_0$
(treated as a point mass particle) of worldline $\mathcal{L}_0$.
Let us recall that, following Einstein's convention for the definition of 
\emph{simultaneity}, the set of events that are considered by $\mathcal{O}_0$
as being simultaneous to a given event $p$ on his worldline is a hypersurface of 
$\M$ which is orthogonal (with respect to $\w{g}$) to $\mathcal{L}_0$ at $p$.

\begin{figure}
\centerline{\includegraphics[height=0.3\textheight]{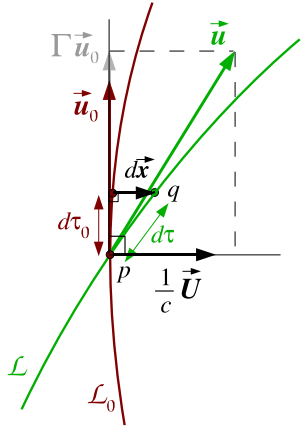}}
\caption{\label{f:lorentz_fact} Relative velocity $\vv{U}=d\vv{x}/d\tau_0$ of
a particle of 4-velocity $\vv{u}$ with respect to an observer of 4-velocity
$\vv{u}_0$. $\vv{U}$ enters in the orthogonal decomposition of $\vv{u}$
with respect to $\vv{u}_0$, via 
$\vv{u} = \Gamma ( \vv{u}_0 + c^{-1}\,  \vv{U} )$.
\emph{NB:} contrary to what the figure might suggest, $d\tau_0 > d\tau$.}
\end{figure}

Let $\mathcal{O}$ be another observer, whose worldline $\mathcal{L}$
intersects that of $\mathcal{O}_0$ at $p$. Let us denote by $\tau_0$ 
(resp. $\tau$) the proper time
of $\mathcal{O}_0$ (resp. $\mathcal{O}$). After some infinitesimal proper time
$d\tau$, $\mathcal{O}$ is located at the point $q$ (cf. Fig.~\ref{f:lorentz_fact}).
Let then $\tau_0+d\tau_0$ be the date attributed by $\mathcal{O}_0$ to the
event $q$ according to the simultaneity convention recalled above. 
The relation between the proper time intervals $d\tau_0$ and
$d\tau$ is
\be \label{e:dtau_Gamma}
	d\tau_0 = \Gamma d\tau , 
\ee
where $\Gamma$ is the \emph{Lorentz factor} between the observers $\mathcal{O}_0$
and $\mathcal{O}$. We can express $\Gamma$ in terms of the 4-velocities
$\vv{u}_0$ and $\vv{u}$ of $\mathcal{O}_0$ and $\mathcal{O}$. Indeed, 
let $d\vv{x}$ the infinitesimal vector that is orthogonal
to $\vv{u}_0$ and links $\mathcal{L}_0$ to $q$ (cf. Fig.~\ref{f:lorentz_fact}).
Since $\vv{u}_0$ and $\vv{u}$ are unit vectors, the following equality holds:
\be \label{e:triangle_u}
	c\, d\tau \vv{u} = c\, d\tau_0 \vv{u}_0 + d\vv{x} . 
\ee
Taking the scalar product with $\vv{u}_0$, and using (\ref{e:dtau_Gamma})
as well as $\vv{u}_0\cdot d\vv{x}=0$ results in 
\be \label{e:Gamma_scal_prod}
	\encadre{ \Gamma = - \vv{u}_0\cdot\vv{u} } .
\ee
Hence from a geometrical point of view, the Lorentz factor is nothing but
(minus) the scalar product of the unit vectors tangent to the two observers' worldlines.

The \emph{velocity of $\mathcal{O}$ relative to} $\mathcal{O}_0$ is simply the
displacement vector $d\vv{x}$ divided by the elapsed proper time of $\mathcal{O}_0$,
$d\tau_0$:
\be
	\vv{U} := \frac{d\vv{x}}{d\tau_0} .
\ee
$\vv{U}$ is the ``ordinary'' velocity, by opposition to the 4-velocity $\vv{u}$.
Contrary to the latter, which is intrinsic to $\mathcal{O}$, $\vv{U}$ depends 
upon the observer $\mathcal{O}_0$.
Geometrically, $\vv{U}$ can be viewed as the part of $\vv{u}$ that is
orthogonal to $\vv{u}_0$, since by combining (\ref{e:dtau_Gamma}) and
(\ref{e:triangle_u}), we get
\be \label{e:decomp_u}
	\encadre{ \vv{u} = \Gamma \left( \vv{u}_0 + \frac{1}{c}\,  \vv{U} \right)},
		\qquad \mbox{with} \quad \vv{u}_0 \cdot \vv{U} = 0 . 
\ee 
Notice that Eq.~(\ref{e:Gamma_scal_prod}) is a mere consequence of the
above relation.
The scalar square of Eq.~(\ref{e:decomp_u}), along with the normalization relations $\vv{u}\cdot\vv{u}=-1$ and $\vv{u}_0\cdot\vv{u}_0=-1$,
leads to
\be
	 \Gamma = \left( 1 - \frac{1}{c^2} \vv{U}\cdot\vv{U} \right) ^{-1/2} , 
\ee
which is identical to the well-known expression from special relativity.


\section{Fluid stress-energy tensor} \label{s:fluid_stress_energy}

\subsection{General definition of the stress-energy tensor}

The \emph{stress-energy tensor} $\w{T}$
is a tensor field on $\M$ which describes the matter
content of spacetime, or more precisely the energy and momentum of matter,
at a macroscopic level. $\w{T}$ is a tensor field of type $(0,2)$ that is
symmetric (this means that at each point $p\in\M$, $\w{T}$ is a symmetric
bilinear form on the vector space $\T_p(\M)$) and that fulfills the 
following properties: given an observer $\mathcal{O}_0$ of 4-velocity
$\vv{u}_0$, 
\begin{itemize}
\item the matter energy density as measured by $\mathcal{O}_0$ is
\be \label{e:T_ener_dens}
	E = \w{T}(\vv{u}_0,\vv{u}_0) ; 
\ee
\item the matter momentum density as measured by $\mathcal{O}_0$ is
\be \label{e:T_mom_dens}
	\vv{p} = -\frac{1}{c} \w{T}(\vv{u}_0, \vv{e}_i)\, \vv{e}_i,
\ee
where $(\vv{e}_i)$ is an orthonormal basis of the hyperplane orthogonal to 
$\mathcal{O}_0$'s worldline (rest frame of $\mathcal{O}_0$);
\item the matter stress tensor as measured by $\mathcal{O}_0$ is
\be \label{e:T_stress}
	S_{ij} = \w{T}(\vv{e}_i, \vv{e}_j) , 
\ee
i.e. $\w{T}(\vv{e}_i, \vv{e}_j)$ is the force in the direction $\vv{e}_i$
acting on the unit surface whose normal is $\vv{e}_j$. 
\end{itemize}

\subsection{Perfect fluid stress-energy tensor}

The \emph{perfect fluid} model of matter relies on a field
of 4-velocities $\vv{u}$, giving at each point the 4-velocity of a fluid
particle. Moreover the perfect fluid is characterized by an isotropic 
pressure in the fluid frame (i.e. $S_{ij} = p\, \delta_{ij}$ for the
observer whose 4-velocity is $\vv{u}$). More precisely, the perfect fluid
model is entirely defined by the following stress-energy tensor:
\be \label{e:T_fluid_parfait}
	\encadre{ \w{T} = (\rho c^2 + p)\,  \uu{u}\otimes\uu{u} + p\, \w{g} }, 
\ee
where $\rho$ and $p$ are two scalar fields, representing respectively
the matter energy density (divided by $c^2$) and the pressure, both measured
in the fluid frame, and $\uu{u}$ is the 1-form associated to the 4-velocity
$\vv{u}$ by the metric tensor $\w{g}$:
\be \label{e:metric_duality}
	\begin{array}{cccl}
	\uu{u} \ : & \T_p(\M) & \longrightarrow & \R \\
		& \vv{v} & \longmapsto & \w{g}(\vv{u},\vv{v}) = \vv{u}\cdot\vv{v}.
	\end{array} 
\ee
In terms of components with respect to a given basis $(\vv{e}_\alpha)$, 
if $\vv{u} = u^\alpha  \vv{e}_\alpha$ and
if $(\w{e}^\alpha)$ is the 1-form basis dual to $(\vv{e}_\alpha)$ 
(cf. Sec.~\ref{s:tensors}), then
$\uu{u} = u_\alpha \w{e}^\alpha$, with $u_\alpha = g_{\alpha\beta} u^\beta$.
In Eq.~(\ref{e:T_fluid_parfait}), the tensor product $\uu{u}\otimes\uu{u}$ stands
for the bilinear form 
$(\vv{v},\vv{w})\mapsto \langle\uu{u},\vv{v}\rangle \langle\uu{u},\vv{w}\rangle=
(\vv{u}\cdot\vv{v})(\vv{u}\cdot\vv{w})$ [cf. (\ref{e:def_tensor_product})].

According to Eq.~(\ref{e:T_ener_dens}) the fluid energy density as
measured by an observer $\mathcal{O}_0$ of 4-velocity $\vv{u}_0$ is 
$E = \w{T}(\vv{u}_0,\vv{u}_0) =
(\rho c^2 + p) (\vv{u}\cdot\vv{u}_0)^2 + p \w{g}(\vv{u}_0,\vv{u}_0)$.
Since $\vv{u}\cdot\vv{u}_0=-\Gamma$, where $\Gamma$ is the Lorentz factor
between the fluid and $\mathcal{O}_0$ [Eq.~(\ref{e:Gamma_scal_prod})],
and $\w{g}(\vv{u}_0,\vv{u}_0) = -1$, we get
\be \label{e:eps_Gamma2}
	E = \Gamma^2 (\rho c^2 + p) - p .
\ee
The reader familiar with the formula $E=\Gamma m c^2$ may be puzzled by the 
$\Gamma^2$ factor in (\ref{e:eps_Gamma2}). However he should remind that 
$E$ is not an energy, but an energy per unit volume: the extra $\Gamma$
factor arises from  ``length contraction'' in the direction of motion.

Similarly, by applying formula (\ref{e:T_mom_dens}), we get the fluid 
momentum density as measured by the observer $\mathcal{O}_0$:
$c \vv{p} = - \w{T}(\vv{u}_0, \vv{e}_i) \, \vv{e}_i
 = - [(\rho c^2 + p) (\vv{u}\cdot\vv{u}_0) (\vv{u}\cdot\vv{e}_i)
	+ p \w{g}(\vv{u}_0,\vv{e}_i) ] \, \vv{e}_i$,
with $\vv{u}\cdot\vv{u}_0=-\Gamma$, $\w{g}(\vv{u}_0,\vv{e}_i) = 0$ and
$(\vv{u}\cdot\vv{e}_i) \, \vv{e}_i$ being the projection of $\vv{u}$ 
orthogonal to $\vv{u}_0$: according to (\ref{e:decomp_u}),
$(\vv{u}\cdot\vv{e}_i) \, \vv{e}_i = \Gamma/c\,  \vv{U}$, where 
$\vv{U}$ is the fluid velocity relative to $\mathcal{O}_0$. Hence
\be \label{e:p_Gamma2}
	\vv{p} = \Gamma^2 \left(\rho + \frac{p}{c^2} \right)  \vv{U} .
\ee

Finally, by applying formula (\ref{e:T_stress}), we get the stress tensor
as measured by the observer $\mathcal{O}_0$:
$S_{ij}= \w{T}(\vv{e}_i, \vv{e}_j)
	= (\rho c^2 + p)(\vv{u}\cdot\vv{e}_i) (\vv{u}\cdot\vv{e}_j) 
	+ p \w{g}(\vv{e}_i,\vv{e}_j)$, 
with $\vv{u}\cdot\vv{e}_i = \Gamma/c\,  \vv{e}_i\cdot\vv{U}$
[thanks to Eq.~(\ref{e:decomp_u}) and $\vv{e}_i\cdot\vv{u}_0=0$] and $\w{g}(\vv{e}_i,\vv{e}_j) = \delta_{ij}$.
Hence
\be \label{e:S_Gamma2}
	S_{ij} =  p \, \delta_{ij} 
		+ \Gamma^2 \left(\rho + \frac{p}{c^2} \right) U^i U^j , 
\ee
where $U^i$ is the $i$-th component of the velocity $\vv{U}$ with 
respect to the orthonormal triad $(\vv{e}_i)$: $\vv{U} = U^i \vv{e}_i$.

Notice that if the observer $\mathcal{O}_0$ is comoving with the fluid, then
$\vv{u}_0 = \vv{u}$, $\Gamma=1$, $\vv{U}=0$ and Eqs.~(\ref{e:eps_Gamma2}),
(\ref{e:p_Gamma2}) and (\ref{e:S_Gamma2}) reduce to 
\be
	E = \rho c^2,\quad
	\vv{p} = 0,\quad S_{ij} =  p \, \delta_{ij} .
\ee
We thus recover the interpretation of the scalar fields $\rho$ and $p$
given above.

\subsection{Concept of equation of state}

Let us assume that at the microscopic level, the perfect fluid is constituted by
$N$ species of particles ($N\geq 1$), so that the energy density $\rho c^2$
is a function of the number densities $n_A$ of particles of species $A$ 
($A\in\{1,2,\ldots,N\}$) in the fluid rest frame (\emph{proper number density})
and of the entropy density $s$ in the fluid rest frame:
\be \label{e:EOS}
	\encadre{ \rho c^2 = \varepsilon(s,n_1,n_2,\ldots,n_N) }.
\ee
The function $\varepsilon$ is called the \emph{equation of state (EOS)} of the fluid.
Notice that $\varepsilon$ is the total energy density, including the rest-mass
energy: denoting by $m^A$ the individual mass of particles of species $A$, 
we may write
\be \label{e:internal_energy}
	\varepsilon = \sum_A m^A n_A c^2 + \varepsilon_{\rm int} , 
\ee
where $\varepsilon_{\rm int}$ is the ``internal'' energy density, 
containing the microscopic kinetic energy of the particles and the 
potential energy resulting from the interactions between the particles.

The first law of thermodynamics in a fixed small comobile volume $V$ writes
\be \label{e:1law0}
	d\mathcal{E} = T \, dS + \sum_A \mu^A \, dN_A , 
\ee
where
\begin{itemize}
\item $\mathcal{E}$ is the total energy in volume $V$: 
$\mathcal{E}=\varepsilon V$,
\item $S$ is the total entropy in $V$: $S=s V$,
\item $N_A$ is the number of particles of species $A$ in $V$: $N_A = n_A V$,
\item $T$ is the \emph{thermodynamical temperature},
\item $\mu^A$ is the \emph{relativistic chemical potential} of particles of
particles of species $A$; it differs from the standard (non-relativistic)
chemical potential ${\tilde\mu}^A$ by the mass $m^A$: 
$\mu^A = {\tilde\mu}^A + m^A$, reflecting the fact that $\mathcal{E}$ includes the
rest-mass energy. 
\end{itemize}
Replacing $\mathcal{E}$, $S$ and $N_A$ by their expression in terms of 
$\varepsilon$, $s$, $n_A$ and $V$ leads to 
$d(\varepsilon V) = T \,d(sV) + \sum_A \mu^A \, d(n_A V)$. Since $V$ is held fixed,
the first law of thermodynamics becomes
\be \label{e:1law}
	\encadre{ d\varepsilon = T \, ds + \sum_A \mu^A \, dn_A }.
\ee
Consequently, $T$ and $\mu^A$ can be expressed as partial derivatives of the
equation of state (\ref{e:EOS}):
\be \label{e:T_muA_partial}
	T = \left( \der{\varepsilon}{s} \right) _{n_A}
	\qquad \mbox{and} \qquad 
	\mu^A = \left( \der{\varepsilon}{n_A} \right) _{s,n_{B\not=A}} .
\ee
Actually these relations can be taken as definitions for the temperature $T$
and chemical potential $\mu^A$. This then leads to the relation (\ref{e:1law}), 
which we will call hereafter the first law of thermodynamics.


\section{Conservation of energy and momentum} \label{s:conser_ener_mom}

\subsection{General form} \label{s:divT_zero}

We shall take for the basis of our presentation of relativistic hydrodynamics
the law of local conservation of energy and momentum of the fluid, which is
assumed to be isolated:
\be \label{e:divT_zero}
	\encadre{ \w{\nabla}\cdot\w{T} = 0 }.
\ee
$\w{\nabla}\cdot\w{T}$ stands for the covariant divergence of the fluid
stress-energy tensor $\w{T}$. This is a 1-form, the components of which in a
given basis are 
\be
	(\w{\nabla}\cdot\w{T})_\alpha = \nabla^\mu T_{\mu\alpha}
	= g^{\mu\nu} \nabla_\nu T_{\mu\alpha} .
\ee
For a self-gravitating fluid, Eq.~(\ref{e:divT_zero}) is actually a consequence 
of the fundamental equation of general relativity, namely the \emph{Einstein equation}. 
Indeed the latter relates the curvature associated with the metric $\w{g}$ to
the matter content of spacetime, according to 
\be \label{e:Einstein}
	\w{G} = \frac{8\pi G}{c^4} \w{T} , 
\ee
where $\w{G}$ is the so-called \emph{Einstein tensor}, which represents some
part of the Riemann curvature tensor of $(\M,\w{g})$. 
A basic property of the Einstein
tensor is $\w{\nabla}\cdot\w{G}=0$ (this follows from the so-called \emph{Bianchi
identities}, which are pure geometric identities regarding the Riemann 
curvature tensor). Thus it is immediate that the Einstein equation
(\ref{e:Einstein}) implies the energy-momentum conservation equation (\ref{e:divT_zero}).
Note that in this respect the situation is different from that of Newtonian theory,
for which the gravitation law (Poisson equation) does not imply the conservation 
of energy and momentum of the matter source.
We refer the reader to \S~22.2 of (Hartle \cite{Hartl03}) for a more extended discussion
of Eq.~(\ref{e:divT_zero}), in particular of the fact that it corresponds only to a \emph{local} conservation of energy and momentum.

Let us mention that there exist formulations of relativistic hydrodynamics
that do not use (\ref{e:divT_zero}) as a starting point, but rather a variational
principle. These Hamiltonian formulations have been pioneered by Taub (\cite{Taub54})
and developed, among others, by Carter (\cite{Carte73}, \cite{Carte79}, \cite{Carte89}), as well as
Comer \& Langlois (\cite{ComerL93}).

\subsection{Application to a perfect fluid}

Substituting the perfect fluid stress-energy tensor (\ref{e:T_fluid_parfait})
in the energy-momentum conservation equation (\ref{e:divT_zero}), and making use
of (\ref{e:nabla_g_zero}) results in 
\be \label{e:divT_fluide_parfait}
\w{\nabla}\cdot\w{T} = 0 \
\iff \   
\left[ \w{\nabla}_{\vv{u}} (\varepsilon + p) + (\varepsilon+p) 
	\w{\nabla}\cdot\vv{u} \right] \uu{u}
	+ (\varepsilon + p) \uu{a} + \w{\nabla} p = 0 , 
\ee
where $\uu{a}$ is the 1-form associated by the metric duality 
[cf. Eq.~(\ref{e:metric_duality}) with $\vv{u}$ replaced by $\vv{a}$]
to the fluid 4-acceleration $\vv{a}=\w{\nabla}_{\vv{u}} \, \vv{u}$
[Eq.~(\ref{e:def_4acc})]. The scalar 
$\w{\nabla}\cdot\vv{u}$ is the covariant divergence
of the 4-velocity vector: it is the trace of the covariant derivative
$\w{\nabla}\vv{u}$, the latter being a type $(1,1)$ tensor: $\w{\nabla}\cdot\vv{u}=\nabla_\sigma u^\sigma$.
Notice that $\w{\nabla} p$ in Eq.~(\ref{e:divT_fluide_parfait}) is nothing but
the gradient of the pressure field: $\w{\nabla} p = \dd p$
(cf. item (i) in Sec.~\ref{s:cov_deriv}).

\subsection{Projection along $\vv{u}$} \label{s:proj_along_u}

Equation (\ref{e:divT_fluide_parfait}) is an identity involving 1-forms.
If we apply it to the vector $\vv{u}$, we get a scalar field.
Taking into account $\langle\uu{u},\vv{u}\rangle = \vv{u}\cdot\vv{u}=-1$ and 
$\langle\uu{a},\vv{u}\rangle=\vv{a}\cdot\vv{u}=0$ [Eq.~(\ref{e:u_ortho_a})], the 
scalar equation becomes
\be \label{e:proj_u_1}
	\langle\w{\nabla}\cdot\w{T},\vv{u}\rangle = 0 \ \iff\ 
	\w{\nabla}_{\vv{u}} \, \varepsilon = - (\varepsilon+p) 
	\w{\nabla}\cdot\vv{u} .
\ee
Notice that $\w{\nabla}_{\vv{u}} \, \varepsilon = \langle\dd \varepsilon,\vv{u}\rangle
= \Lie{u}\varepsilon$ (cf. item (i) at the end of Sec.~\ref{s:lie_deriv}).
Now, the first law of thermodynamics (\ref{e:1law}) yields
\be
	\w{\nabla}_{\vv{u}} \, \varepsilon = T \w{\nabla}_{\vv{u}} \, s
	+ \sum_A \mu^A \w{\nabla}_{\vv{u}} \, n_A ,
\ee
so that Eq.~(\ref{e:proj_u_1}) can be written as
\bea
	\langle\w{\nabla}\cdot\w{T},\vv{u}\rangle = 0 & \iff & 
	T \w{\nabla}\cdot(s\vv{u}) + \sum_A \mu^A \w{\nabla}\cdot(n_A \vv{u}) 
		\nonumber \\
	& & \ \quad + \left( \varepsilon + p - T s - \sum_A \mu^A n_A   \right)
	\w{\nabla}\cdot\vv{u} = 0. \label{e:proj_u_2}
\eea
Now, we recognize in $\mathcal{G}:= \varepsilon + p - T s$ the \emph{free enthalpy}
(also called \emph{Gibbs free energy}) per unit volume. It is well known that the
free enthalpy $G=\mathcal{G} V = E + PV -TS$ (where $V$ is some small volume element) obeys the thermodynamic identity
\be
	G = \sum_A \mu^A N_A , 
\ee
from which we get $\mathcal{G} = \sum_A \mu^A n_A$, i.e.
\be \label{e:p_TS_etc}
	\encadre{ p = T s + \sum_A \mu^A n_A - \varepsilon }.
\ee
This relation shows that $p$ is a function of $(s,n_1,\ldots,n_N)$ which is 
fully determined by $\varepsilon(s,n_1,\ldots,n_N)$ [recall that $T$ and $\mu^A$
are nothing but partial derivatives of the latter, Eq.~(\ref{e:T_muA_partial})].
Another way to get the identity (\ref{e:p_TS_etc}) is to start from the first law
of thermodynamics in the form (\ref{e:1law0}), but allowing for the volume $V$
to vary, i.e. adding the term $-p\,  dV$ to it:
\be 
	d\mathcal{E} = T \, dS - p\,  dV + \sum_A \mu^A \, dN_A .
\ee
Substituting $\mathcal{E}=\varepsilon V$, $S=s V$ and $N_A = n_A V$ is this formula
and using (\ref{e:1law}) leads to (\ref{e:p_TS_etc}).

For our purpose the major consequence of the thermodynamic identity 
(\ref{e:p_TS_etc}) is that Eq.~(\ref{e:proj_u_2}) simplifies substantially:
\be \label{e:proj_u_3}
	\langle\w{\nabla}\cdot\w{T},\vv{u}\rangle = 0 \  \iff \
	\encadre{ T \w{\nabla}\cdot(s\vv{u}) + \sum_A \mu^A \w{\nabla}\cdot(n_A \vv{u}) 
	= 0 }.
\ee
In this equation, $c \w{\nabla}\cdot(s\vv{u})$ is the entropy creation rate
(entropy created per unit volume and unit time in the fluid frame)
and $c \w{\nabla}\cdot(n_A \vv{u})$ is the particle creation rate of species $A$
(number of particles created par unit volume and unit time in the fluid frame).
This follows from 
\be
	c \w{\nabla}\cdot(n_A \vv{u}) = c(\w{\nabla}_{\vv{u}} n_A 
	+ n_A \w{\nabla}\cdot\vv{u}) = \frac{dn_A}{d\tau} 
	+ n_A \frac{1}{V} \frac{dV}{d\tau}
	= \frac{1}{V} \frac{d(n_A V)}{d\tau} , 
\ee
where $\tau$ is the fluid proper time and where we have used the expansion rate
formula
\be \label{e:expansion_rate}
	\w{\nabla}\cdot\vv{u} = \frac{1}{c V} \frac{dV}{d\tau} , 
\ee
$V$ being a small volume element dragged along by $\vv{u}$.
Equation (\ref{e:proj_u_3}) means that in a perfect fluid, the only process that
may increase the entropy is the creation of particles.

\subsection{Projection orthogonally to $\vv{u}$: relativistic Euler equation}

Let us now consider the projection of (\ref{e:divT_fluide_parfait})
orthogonally to the 4-velocity. The projector orthogonal to $\vv{u}$ is
the operator $\w{P} := \w{1} + \vv{u}\otimes \uu{u}$:
\be
	\begin{array}{cccl}
	\w{P} \ : & \T_p(\M) & \longrightarrow &   \T_p(\M) \\
		& \vv{v} & \longmapsto & \vv{v} + (\vv{u}\cdot\vv{v}) \vv{u} . 
	\end{array} 
\ee
Combining $\w{P}$ to the 1-form (\ref{e:divT_fluide_parfait}), and using
$\uu{u}\circ\w{P}=0$ as well as $\uu{a}\circ\w{P}=\uu{a}$, leads to the 1-form equation
\be \label{e:Euler_relat_u}
(\w{\nabla}\cdot\w{T})\circ\w{P} = 0 \
\iff \   
	\encadre{ (\varepsilon + p) \uu{a} = - \w{\nabla} p 
	- (\w{\nabla}_{\vv{u}}\, p) \uu{u} } .
\ee
This is clearly an equation of the type `` $m \vec{a} = \vec{F}$ '', although
the gravitational ``force'' is hidden in the covariant derivative
in the derivation of $\vv{a}$ from $\vv{u}$. We may consider that (\ref{e:Euler_relat_u})
is a relativistic version of the classical Euler equation.

Most textbooks stop at this point, considering that (\ref{e:Euler_relat_u})
is a nice equation. However, as stated in the Introduction, there exists an alternative form for the equation of motion
of a perfect fluid, which turns out to be much more useful than (\ref{e:Euler_relat_u}), 
especially regarding the derivation of conservation laws: it is the Carter-Lichnerowicz
form, to which the rest of this lecture is devoted.


\section{Carter-Lichnerowicz equation of motion}  \label{s:CL_eom}

\subsection{Derivation}

In the right hand-side of the relativistic Euler equation (\ref{e:Euler_relat_u}) 
appears the gradient of the pressure field: $\w{\nabla} p = \dd p$. 
Now, by deriving the thermodynamic identity (\ref{e:p_TS_etc}) and combining with 
the first law (\ref{e:1law}), we get the relation
\be
	\encadre{ dp = s \, dT + \sum_A n_A d\mu^A }, 
\ee
which is known as the \emph{Gibbs-Duhem relation}.
We may use this relation to express $\w{\nabla} p$ in terms of $\w{\nabla} T$
and $\w{\nabla} \mu^A$ in Eq.~(\ref{e:Euler_relat_u}). Also, by making use
of (\ref{e:p_TS_etc}), we may replace $\varepsilon+p$ by $T s + \sum_A \mu^A n_A$.
Hence Eq.~(\ref{e:Euler_relat_u}) becomes
\be
	\left(T s +  \sum_A \mu^A n_A\right) \uu{a} = 
	- s \w{\nabla} T - \sum_A n_A \w{\nabla} \mu^A 
	- \left(  s \w{\nabla}_{\vv{u}}\, T + \sum_A n_A \w{\nabla}_{\vv{u}}\,  \mu^A
	\right) \uu{u} .  
\ee
Writing $\uu{a} = \vv{\nabla}_{\vv{u}} \uu{u}$ and reorganizing slightly yields
\be \label{e:CL_prov}
	s \left[ \w{\nabla}_{\vv{u}}(T\uu{u}) + \w{\nabla} T \right]
	+ \sum_A n_A \left[  \w{\nabla}_{\vv{u}}(\mu^A\uu{u})
	+ \w{\nabla} \mu^A \right] = 0 . 
\ee
The next step amounts to noticing that 
\be \label{e:nabu_Lieu}
	\w{\nabla}_{\vv{u}}(T\uu{u}) = \Lie{u} (T \uu{u}) . 
\ee
This is easy to establish, starting from expression (\ref{e:Lie_der_1form}) for
the Lie derivative of a 1-form, in which we may replace the partial
derivatives by covariant derivatives [thanks to the symmetry of the Christoffel
symbols, cf. Eq.~(\ref{e:cov_derivT_comp})]:
\be
	\Liec{u} (T u_\alpha) = u^\mu \nabla_\mu (T u_\alpha)
	+ T u_\mu \nabla_\alpha u^\mu ,
\ee
i.e.
\be
	\Lie{u} (T \uu{u}) = \w{\nabla}_{\vv{u}}(T\uu{u}) + T\,  \vv{u}\cdot
	\w{\nabla} \vv{u} . 
\ee
Now, from $\vv{u}\cdot\vv{u}=-1$, we get $\vv{u}\cdot\w{\nabla} \vv{u}=0$,
which establishes (\ref{e:nabu_Lieu}).

On the other side, the Cartan identity (\ref{e:Cartan}) yields
\be
	\Lie{u} (T \uu{u}) = \vv{u}\cdot\dd (T \uu{u})
	+ \dd[ T\underbrace{\langle\uu{u},\vv{u}\rangle}_{=-1}]
	= \vv{u}\cdot\dd (T \uu{u}) - \dd T .
\ee
Combining this relation with (\ref{e:nabu_Lieu}) (noticing that
$\dd T = \w{\nabla} T$), we get 
\be
	\w{\nabla}_{\vv{u}}(T\uu{u}) + \w{\nabla} T =  \vv{u}\cdot\dd (T \uu{u}) .
\ee
Similarly, 
\be
	\w{\nabla}_{\vv{u}}(\mu^A\uu{u}) + \w{\nabla} \mu^A = \vv{u}\cdot
	\dd (\mu^A \uu{u}) .
\ee
According to the above two relations, the equation of motion (\ref{e:CL_prov}) can be re-written as 
\be \label{e:CL_prov2}
	s \, \vv{u}\cdot\dd (T \uu{u}) 
	+ \sum_A n_A \, \vv{u}\cdot\dd (\mu^A \uu{u}) = 0 .
\ee
In this equation, appears the 1-form
\be \label{e:def_piA}
	\encadre{ \w{\pi}^A := \mu^A \uu{u} },
\ee
which is called the \emph{momentum 1-form} of particles of species $A$.
It is called \emph{momentum} because in the Hamiltonian formulations mentioned
in Sec.~\ref{s:divT_zero}, this 1-form is the conjugate of the number density
current $n_A \vv{u}$.

Actually, it is the exterior derivative of $\w{\pi}^A$ which appears in 
Eq.~(\ref{e:CL_prov2}):
\be \label{e:def_wA}
	\encadre{ \w{w}^A := \dd \w{\pi}^A }.
\ee
This 2-form is called the \emph{vorticity 2-form} of particles of species $A$.
With this definition, Eq.~(\ref{e:CL_prov2}) becomes
\be \label{e:CL}
	\encadre{ \sum_A n_A \, \vv{u}\cdot\w{w}^A
	+ s \, \vv{u}\cdot\dd (T \uu{u}) = 0 }.
\ee
This is the \emph{Carter-Lichnerowicz form} of the equation of motion for a
multi-constituent perfect fluid.
It has been considered by Lichnerowicz (\cite{Lichn67}) in the case of a single-constituent 
fluid ($N=1$) and generalized by Carter (\cite{Carte79}, \cite{Carte89}) to the multi-constituent 
case.
Let us stress that this is an equation between 1-forms. For instance
$\vv{u}\cdot\w{w}^A$ is the 1-form $\w{w}^A(\vv{u},.)$, i.e. at each point 
$p\in\M$, this is the linear application $\T_p \rightarrow\R$, 
$\vv{v}\mapsto \w{w}^A(\vv{u},\vv{v})$. Since $\w{w}^A$ is antisymmetric (being a
2-form), $\w{w}^A(\vv{u},\vv{u})=0$. Hence the Carter-Lichnerowicz equation 
(\ref{e:CL}) is clearly a non-trivial equation only in the three dimensions 
orthogonal to $\vv{u}$.

\subsection{Canonical form for a simple fluid} \label{s:simple_fluid}

Let us define a \emph{simple fluid} as a fluid for which the EOS (\ref{e:EOS})
takes the form
\be \label{e:EOS_sf}
	\encadre{ \varepsilon = \varepsilon(s,n) }, 
\ee
where $n$ is the baryon number density in the fluid rest frame.
The simple fluid model is valid in two extreme cases:
\begin{itemize}
\item when the reaction rates between the various particle species are very low:
the composition of matter is then frozen: all the particle number densities
can be deduced from the baryon number: $n_A = Y_A n$, with a fixed species fraction
$Y_A$;
\item when reaction rates between the various particle species are very high,
ensuring a complete chemical (nuclear) equilibrium. In this case, all the $n_A$
are uniquely determined by $n$ and $s$, via $n_A = Y_A^{\rm eq}(s,n)\, n$.
\end{itemize}

A special case of a simple fluid is that of \emph{barotropic fluid}, for which
\be \label{e:EOS_baro}
	\varepsilon = \varepsilon(n) .
\ee
This subcase is particularly relevant for cold dense matter, as in white dwarfs and
neutron stars.

Thanks to Eq.~(\ref{e:EOS_sf}), a simple fluid behaves as if it contains a single
particle species: the baryons. All the equations derived previously apply,
setting $N=1$ (one species) and $A=1$. 

Since $n$ is the baryon number density, it must obey the fundamental law
of \emph{baryon conservation}:
\be  \label{e:baryon_cons}
	\encadre{ \w{\nabla}\cdot(n \vv{u}) = 0 }.
\ee
That this equation does express the conservation of baryon number should be
obvious after the discussion in Sec.~\ref{s:proj_along_u}, from which it follows
that $c \w{\nabla}\cdot(n \vv{u})$ is the number of baryons created per unit volume
and unit time in a comoving fluid element.

The projection of $\w{\nabla}\cdot\w{T}=0$ along $\vv{u}$, Eq.~(\ref{e:proj_u_3})
then implies
\be \label{e:entropy_cons}
	\encadre{ \w{\nabla}\cdot(s \vv{u}) = 0 }.
\ee
This means that the evolution of a (isolated) simple fluid is necessarily adiabatic.

On the other side, the Carter-Lichnerowicz equation (\ref{e:CL}) reduces to
\be  \label{e:CL_sf_prov}
	n \, \vv{u}\cdot\w{w}^{\rm b}
	+ s \, \vv{u}\cdot\dd (T \uu{u}) = 0 ,
\ee
where we have used the label ${\rm b}$ (for baryon) instead of the running letter $A$:
$\w{w}^{\rm b} = \dd(\mu \uu{u})$ [Eqs.~(\ref{e:def_piA}) and (\ref{e:def_wA})],
$\mu$ being the chemical potentials of baryons:
\be
	\mu := \left( \der{\varepsilon}{n} \right) _s .
\ee
Let us rewrite Eq.~(\ref{e:CL_sf_prov}) as
\be \label{e:CL_sf_prov2}
	\vv{u} \cdot \left[ \dd (\mu \uu{u}) + {\bar s}\,  \dd(T \uu{u}) \right] = 0 ,
\ee
where we have introduced the \emph{entropy per baryon}:
\be
	\encadre{ \bar s := \frac{s}{n} }.
\ee

In view of Eq.~(\ref{e:CL_sf_prov2}), let us define the \emph{fluid momentum 
per baryon 1-form} by
\be \label{e:def_pi_mu_Ts}
	\w{\pi} := (\mu + T {\bar s} ) \uu{u} 
\ee
and the \emph{fluid vorticity 2-form} as its exterior derivative:
\be \label{e:def_vorticity}
	\encadre{ \w{w} := \dd \w{\pi} }.
\ee
Notice that $\w{w}$ is not equal to the baryon vorticity:
$\w{w} = \w{w}^{\rm b} + \dd({\bar s} T \uu{u})$.
Since in the present case the thermodynamic identity (\ref{e:p_TS_etc}) reduces
to $p = T s + \mu n - \varepsilon$, we have
\be \label{e:def_enthalpy_bar}
	\encadre{ \mu + T {\bar s} = \frac{\varepsilon + p}{n} =: h },
\ee
where $h$ is the \emph{enthalpy per baryon}.
Accordingly the fluid momentum per baryon 1-form is simply
\be \label{e:pi_hu}
	\encadre{ \w{\pi} = h\, \uu{u} }. 
\ee
By means of formula (\ref{e:def_ext_1f}), we can expand the exterior derivative
$\dd({\bar s} T \uu{u})$ as 
\be \label{e:expand_dsTu}
	\dd({\bar s} T \uu{u}) = \dd {\bar s} \wedge (T\uu{u})
	+ {\bar s} \, \dd(T\uu{u}) 
	= T \dd {\bar s} \wedge \uu{u}
	+ {\bar s} \, \dd(T\uu{u}) ,
\ee
where the symbol $\wedge$ stands for the \emph{exterior product}: for any 
pair $(\w{a},\w{b})$ of 1-forms, $\w{a}\wedge\w{b}$ is the 2-form defined
as $\w{a}\wedge\w{b} = \w{a}\otimes\w{b} - \w{b}\otimes\w{a}$.
Thanks to (\ref{e:expand_dsTu}), the fluid vorticity 2-form, 
given by Eqs.~(\ref{e:def_vorticity}) and (\ref{e:def_pi_mu_Ts}), can be
written as  
\be
	\w{w} = \dd(\mu \uu{u}) + {\bar s} \, \dd(T\uu{u})
	+ T \dd {\bar s} \wedge \uu{u} .
\ee
Therefore, we may rewrite (\ref{e:CL_sf_prov2}) by letting appear $\w{w}$,
to get successively
\bea
	& & \vv{u}\cdot (\w{w} - T \dd {\bar s} \wedge \uu{u}) = 0 ,\nonumber \\
	& & \vv{u}\cdot \w{w} - T \vv{u} \cdot ( \dd {\bar s} \wedge \uu{u}) = 0 ,	
		\nonumber \\
	& & \vv{u}\cdot \w{w} - T \left[ \langle\dd{\bar s},\vv{u}\rangle\, \uu{u}
	-\langle\uu{u},\vv{u}\rangle \dd{\bar s} \right] = 0 , \nonumber \\
	& & \vv{u}\cdot \w{w} - T \left[ (\w{\nabla}_{\vv{u}}\, {\bar s}) \, \uu{u}
	+  \dd{\bar s} \right] = 0 . \label{e:CL_sf_prov3}
\eea
Now from the baryon number and entropy conservation equations (\ref{e:baryon_cons})
and (\ref{e:entropy_cons}), we get
\be
	\encadre{ \w{\nabla}_{\vv{u}}\, {\bar s} = 0 },
\ee
i.e. the entropy per baryon is conserved along the fluid lines.
Reporting this property in Eq.~(\ref{e:CL_sf_prov3}) leads to the
equation of motion
\be \label{e:CL_sf}
	\encadre{ \encadre{ \vv{u}\cdot \w{w} = T \dd {\bar s} }}.
\ee
This equation was first obtained by Lichnerowicz (\cite{Lichn67}).
In the equivalent form (assuming $T\not =0$),
\be
	\vv{u}'\cdot \w{w} = \dd {\bar s} ,\qquad \mbox{with}\quad
	\vv{u}' := \frac{1}{T} \vv{u} , 
\ee
it has been called a \emph{canonical} equation of motion by Carter (\cite{Carte79}),
who has shown that it can be derived from a variational principle.

Owing to its importance, let us make (\ref{e:CL_sf}) explicit in terms of
components [cf. (\ref{e:def_ext_0f}) and (\ref{e:def_ext_1f})]:
\be \label{e:CL_sf_comp}
	u^\mu \left[ \der{}{x^\mu}(h \, u_\alpha) 
	- \der{}{x^\alpha}(h\, u_\mu) \right] = T \der{{\bar s}}{x^\alpha} .
\ee

\subsection{Isentropic case (barotropic fluid)} \label{s:isentropic}

For an isentropic fluid, ${\bar s}={\rm const}$.
The EOS is then barotropic, i.e. it can be cast in the form (\ref{e:EOS_baro}).
For this reason, the isentropic simple fluid is also called a \emph{single-constituent
fluid}. In this case, the gradient $\dd {\bar s}$ vanishes and the Carter-Lichnerowicz equation of motion (\ref{e:CL_sf}) reduces to
\be \label{e:CL_baro}
	\encadre{ \vv{u}\cdot \w{w} = 0 }.
\ee
This equation has been first exhibited by Synge (\cite{Synge37}).
Its simplicity  is remarkable, especially if we compare it
to the equivalent Euler form (\ref{e:Euler_relat_u}). Indeed it should be noticed
that the assumption of a single-constituent fluid leaves 
the relativistic Euler equation as it is written in (\ref{e:Euler_relat_u}), whereas it 
leads to the simple form (\ref{e:CL_baro}) for the Carter-Lichnerowicz 
equation of motion.

In the isentropic case, there is a useful relation between the gradient of pressure
and that of the enthalpy per baryon. Indeed, from  
Eq.~(\ref{e:def_enthalpy_bar}), we have 
$d\varepsilon + dp = d(n h) = n \, dh + h \, dn$. Substituting Eq.~(\ref{e:1law})
for $d\varepsilon$ yields $T\, ds + \mu \, dn + dp = n \, dh + h \, dn$.
But $T\, ds = T\, d(n {\bar s}) = T{\bar s}\, dn$ since $d{\bar s}=0$.
Using Eq.~(\ref{e:def_enthalpy_bar}) again then leads to 
\be
	dp = n \, dh ,
\ee
or equivalently,
\be
 \frac{dp}{\varepsilon+p} = d\ln h .
\ee
If we come back to the relativistic Euler equation (\ref{e:Euler_relat_u}), 
the above
relation shows that in the isentropic case, it can be written as the fluid
4-acceleration being the orthogonal projection (with respect to $\vv{u}$) of
a pure gradient (that of $-\ln h$):
\be
	\uu{a} = - \dd\ln h - \langle\dd\ln h,\vv{u}\rangle \, \uu{u} .
\ee

\subsection{Newtonian limit: Crocco equation} \label{s:Crocco}

Let us go back to the non isentropic case and consider the
Newtonian limit of the Carter-Lichnerowicz equation (\ref{e:CL_sf}). 
For this purpose let us assume that the gravitational field is weak and static.
It is then always possible to find a coordinate system $(x^\alpha)=(x^0=ct,x^i)$ such that the metric components take the form
\be \label{e:gab_weak}
   g_{\alpha\beta} dx^\alpha dx^\beta =
	- \left( 1 + 2\frac{\Phi}{c^2} \right)  \, c^2 dt^2
	+ \left( 1 -2 \frac{\Phi}{c^2} \right) f_{ij} dx^i dx^j,
\ee
where $\Phi$ is the Newtonian gravitational potential (solution of
$\Delta\Phi=4\pi G\rho$) and $f_{ij}$ is the flat metric in the usual 3-dimensional
Euclidean space. For a weak gravitational field (Newtonian limit), 
$|\Phi|/c^2 \ll 1$.
The components of the fluid 4-velocity are deduced from Eq.~(\ref{e:def_4vel}):
$u^\alpha = c^{-1} dx^\alpha/d\tau$, $\tau$ being the fluid proper time.
Thus (recall that $x^0=ct$)
\be \label{e:ua_Newt_lim_prov}
	u^\alpha = \left(u^0, u^0 \frac{v^i}{c} \right), \quad
	\mbox{with} \quad u^0 = \frac{dt}{d\tau} \quad\mbox{and}\quad
	v^i := \frac{dx^i}{dt} .
\ee
At the Newtonian limit, the $v^i$'s are of course the components of the fluid velocity
$\vec{v}$ with respect to the inertial frame defined by the coordinates $(x^\alpha)$.
That the coordinates $(x^\alpha)$ are inertial in the Newtonian limit is obvious from
the form (\ref{e:gab_weak}) of the metric, which is clearly Minkowskian when
$\Phi\rightarrow 0$. Consistent with the Newtonian limit, we assume that
$|\vec{v}|/c\ll 1$. The normalization relation $g_{\alpha\beta} u^\alpha u^\beta = -1$
along with (\ref{e:ua_Newt_lim_prov})
enables us to express $u^0$ in terms of $\Phi$ and $\vec{v}$. To the first order
in $\Phi/c^2$ and $\vec{v}\cdot\vec{v}/c^2=v_j v^j /c^2$ (\footnote{the indices of
$v^i$ are lowered by the flat metric: $v_i := f_{ij} v^j$}), we get
\be \label{e:u0_Newt_lim}
	u^0 \simeq 1 - \frac{\Phi}{c^2}  + \frac{v_j v^j}{2 c^2} .
\ee
To that order of approximation, we may set $u^0\simeq 1$ in the spatial part of
$u^\alpha$ and rewrite (\ref{e:ua_Newt_lim_prov}) as
\be \label{e:ua_Newt_lim}
	u^\alpha \simeq \left( u^0, \frac{v^i}{c} \right) 
	\simeq \left( 1 - \frac{\Phi}{c^2}  + \frac{v_j v^j}{2 c^2} ,\ 
	\frac{v^i}{c} \right) .
\ee
The components of $\uu{u}$ are obtained from $u_\alpha = g_{\alpha\beta} u^\beta$,
with $g_{\alpha\beta}$ given by (\ref{e:gab_weak}). One gets
\be \label{e:ucova_Newt_lim}
	u_\alpha \simeq \left( u_0, \frac{v_i}{c} \right) \simeq
	\left( -1 - \frac{\Phi}{c^2}  - \frac{v_j v^j}{2 c^2} ,\  \frac{v_i}{c} \right) .
\ee

To form the fluid vorticity $\w{w}$ we need the enthalpy per baryon $h$.
By combining Eq.~(\ref{e:def_enthalpy_bar}) with Eq.~(\ref{e:internal_energy})
written as $\varepsilon = m_{\rm b} n c^2 + \varepsilon_{\rm int}$ (where 
$m_{\rm b}$ is the mean mass of one baryon: $m_{\rm b}\simeq 1.66\times 10^{-27}{\rm\ kg}$), we get
\be \label{e:h_Newt_lim}
	h = m_{\rm b} c^2 \left( 1 + \frac{H}{c^2} \right),
\ee
where $H$ is the non-relativistic (i.e. excluding the rest-mass energy) 
\emph{specific enthalpy} (i.e. enthalpy per unit mass):
\be
	H := \frac{\varepsilon_{\rm int} + p}{m_{\rm b} n} .
\ee
From (\ref{e:CL_sf_comp}), we have, for $i\in\{1,2,3\}$,
\bea
	u^\mu w_{\mu i} & = & u^\mu \left[ \der{}{x^\mu}(h \, u_i) 
	- \der{}{x^i}(h\, u_\mu) \right] \nonumber \\
	& = & u^0  \left[ \frac{1}{c}\der{}{t}(h \, u_i) 
	- \der{}{x^i}(h\, u_0) \right]
	+ u^j \left[ \der{}{x^j}(h \, u_i) 
	- \der{}{x^i}(h\, u_j) \right] .
\eea
Plugging Eqs.~(\ref{e:ua_Newt_lim}), (\ref{e:ucova_Newt_lim}) and (\ref{e:h_Newt_lim})
yields
\bea
	\frac{u^\mu w_{\mu i}}{m_{\rm b}} & = &
	u^0 \left\{ \der{}{t} \left[ \left(1+\frac{H}{c^2}\right) v_i \right]
	-  \der{}{x^i} \left[ (c^2 + H) u_0 \right] \right\} \nonumber \\
	& & + v^j \left\{  \der{}{x^j}\left[ \left(1+\frac{H}{c^2}\right) v_i \right]
	- \der{}{x^i} \left[ \left(1+\frac{H}{c^2}\right) v_j \right] \right\} .
\eea
At the Newtonian limit, the terms $u^0$ and $H/c^2$ in the above equation can 
be set to respectively $1$ and $0$. Moreover, thanks to (\ref{e:ucova_Newt_lim}),
\bea
    \der{}{x^i} \left[ (c^2 + H) u_0 \right] & = &
	- \der{}{x^i} \left[ (c^2 + H) \left( 1 + 
	\frac{\Phi}{c^2} + \frac{v_j v^j}{2 c^2} \right) \right] \nonumber \\
	& \simeq & - \der{}{x^i} \left( \Phi + \frac{1}{2} v_j v^j
	 + H \right) .
\eea
Finally we get 
\be \label{e:CL_Newt_lim_prov}
	\frac{u^\mu w_{\mu i}}{m_{\rm b}}  = \der{v_i}{t}
	+ \der{}{x^i} \left( H + \frac{1}{2} v_j v^j + \Phi \right)
	+ v^j \left( \der{v_i}{x^j} - \der{v_j}{x^i} \right) .
\ee
The last term can be expressed in terms of the cross product between
$\vec{v}$ and its the (3-dimensional) curl:
\be \label{e:curl_v}
	v^j \left( \der{v_i}{x^j} - \der{v_j}{x^i} \right) = 
	- \left( \vec{v} \times \mathrm{curl} \, \vec{v} \right)_i .
\ee
In view of (\ref{e:CL_Newt_lim_prov}) and (\ref{e:curl_v}), we conclude that 
the Newtonian limit of the Carter-Lichnerowicz canonical equation
(\ref{e:CL_sf}) is
\be \label{e:Crocco}
	\der{v_i}{t} + \der{}{x^i} \left( H + \frac{1}{2} v_j v^j + \Phi \right)
	- \left( \vec{v} \times \mathrm{curl} \, \vec{v} \right)_i
	= T \der{{\tilde s}}{x^i} ,
\ee
where ${\tilde s} := {\bar s} / m_{\rm b}$ is the 
\emph{specific entropy} (i.e. entropy per unit mass).
Equation (\ref{e:Crocco}) is known as the \emph{Crocco equation} [see e.g. (Rieutord
\cite{Rieut97})]. It is of course an alternative form of the classical 
Euler equation in the gravitational potential $\Phi$.


\section{Conservation theorems} \label{s:conserv_theor}

In this section, we illustrate the power of the Carter-Lichnerowicz equation
by deriving from it various conservation laws in a very easy way.
We consider a simple fluid, i.e. the EOS depends only on the baryon
number density and the entropy density [Eq.~(\ref{e:EOS_sf})].

\subsection{Relativistic Bernoulli theorem}

\subsubsection{Conserved quantity associated with a spacetime symmetry}

Let us suppose that the spacetime $(\M,\w{g})$ has some symmetry described by 
the invariance under the action of a one-parameter group $\mathcal{G}$:
for instance $\mathcal{G}=(\R,+)$ for stationarity (invariance by translation along
timelike curves) or $\mathcal{G}=\mathrm{SO(2)}$ for axisymmetry 
(invariance by rotation around some axis). Then one can associate to $\mathcal{G}$
a vector field $\vv{\xi}$ such that an infinitesimal transformation of parameter $\epsilon$ in the group $\mathcal{G}$ corresponds to the infinitesimal displacement $\epsilon\vv{\xi}$. In particular the field lines of 
$\vv{\xi}$ are the trajectories (also called orbits) of $\mathcal{G}$.
$\vv{\xi}$ is called a \emph{generator of the symmetry group} $\mathcal{G}$ or
a \emph{Killing vector} of spacetime.
That the metric tensor $\w{g}$ remains invariant under $\mathcal{G}$ is then
expressed by the vanishing of the Lie derivative of $\w{g}$ along $\vv{\xi}$:
\be \label{e:Killing}
	\encadre{ \Lie{\xi} \w{g} = 0 }.
\ee
Expressing the Lie derivative via Eq.~(\ref{e:Lie_der_comp}) with the partial 
derivatives replaced by  covariant ones [cf. remark below Eq.~(\ref{e:nabu_Lieu})], we immediately get that 
(\ref{e:Killing}) is equivalent to the following requirement on the 1-form
$\uu{\xi}$ associated to $\vv{\xi}$ by the metric duality:
\be \label{e:Killing_eq}
	\nabla_\alpha \xi_\beta + \nabla_\beta \xi_\alpha = 0 .
\ee
Equation (\ref{e:Killing_eq}) is called the \emph{Killing equation}. It fully 
characterizes Killing vectors in a given spacetime.

The invariance of the fluid under the symmetry group $\mathcal{G}$
amounts to the vanishing of the Lie derivative along $\vv{\xi}$
of all the tensor fields associated with matter. In particular, for the
fluid momentum per baryon 1-form $\w{\pi}$ introduced in Sec.~\ref{s:simple_fluid}:
\be
	\Lie{\xi} \w{\pi} = 0 .
\ee
By means of the Cartan identity (\ref{e:Cartan}), this equation is recast as
\be \label{e:xi_w}
	\vv{\xi}\cdot\w{w} + \dd \langle\w{\pi},\vv{\xi}\rangle  = 0 ,
\ee
where we have replaced the exterior derivative $\dd\w{\pi}$ by the vorticity
2-form $\w{w}$ [cf. Eq.~(\ref{e:def_vorticity})] and we have written
$\vv{\xi}\cdot\w{\pi}=\langle\w{\pi},\vv{\xi}\rangle$ 
(scalar field resulting from the action
of the 1-form $\w{\pi}$ on the vector $\vv{\xi}$).
The left-hand side of Eq.~(\ref{e:xi_w}) is a 1-form. Let us apply it to the vector
$\vv{u}$:
\be
	\w{w}(\vv{\xi},\vv{u}) + \w{\nabla}_{\vv{u}} \langle\w{\pi},\vv{\xi}\rangle = 0  
\ee
Now, since $\w{w}$ is antisymmetric, $\w{w}(\vv{\xi},\vv{u}) = - \w{w}(\vv{u},\vv{\xi})$
and we may use the Carter-Lichnerowicz equation of motion (\ref{e:CL_sf}) which
involves $\w{w}(\vv{u},.) = \vv{u}\cdot\w{w}$ to get
\be
	- T \langle\dd{\bar s},\vv{\xi}\rangle + \w{\nabla}_{\vv{u}} 
	\langle\w{\pi},\vv{\xi}\rangle = 0 .
\ee
But $\langle\dd{\bar s}, \vv{\xi}\rangle = \Lie{\xi} {\bar s}$
and, by the fluid symmetry under $\mathcal{G}$, $\Lie{\xi} {\bar s} = 0$.
Therefore there remains
\be
	\w{\nabla}_{\vv{u}} \langle\w{\pi},\vv{\xi}\rangle = 0 ,
\ee
which, thanks to Eq.~(\ref{e:pi_hu}), we may rewrite as
\be \label{e:Bernoul_gal}
	\encadre{ \w{\nabla}_{\vv{u}} (h\, \vv{\xi}\cdot\vv{u}) = 0 }.
\ee
We thus have established that if $\vv{\xi}$ is a symmetry generator of spacetime,
the scalar field $h\, \vv{\xi}\cdot\vv{u}$ remains constant along the flow lines.

The reader with a basic knowledge of relativity must have noticed the similarity
with the existence of conserved quantities along the geodesics in symmetric 
spacetimes: if $\vv{\xi}$ is a Killing vector, it is well known that 
the quantity $\vv{\xi}\cdot\vv{u}$ is conserved along any timelike geodesic
($\vv{u}$ being the 4-velocity associated with the geodesic) 
[see e.g. Chap.~8 of (Hartle \cite{Hartl03})]. In the present case, it is not
the quantity $\vv{\xi}\cdot\vv{u}$ which is conserved along the flow lines
but $h\, \vv{\xi}\cdot\vv{u}$. The ``correcting factor'' $h$ arises because the 
fluid worldlines are not geodesics due to the pressure in the fluid. 
As shown by the relativistic Euler equation (\ref{e:Euler_relat_u}), they are 
geodesics ($\uu{a}=0$) only if $p$ is constant (for instance $p=0$).

\subsubsection{Stationary case: relativistic Bernoulli theorem}

In the case where the Killing vector $\vv{\xi}$ is timelike, the spacetime is
said to be \emph{stationary} and Eq.~(\ref{e:Bernoul_gal}) 
constitutes the relativistic generalization of the classical \emph{Bernoulli theorem}. 
It was first established by Lichnerowicz (\cite{Lichn40}) (see also Lichnerowicz
\cite{Lichn41}), the special relativistic
subcase (flat spacetime) having been obtained previously by Synge (\cite{Synge37}).

By means of the formul\ae\ established in Sec.~\ref{s:Crocco}, it is easy
to see that at the Newtonian limit, Eq.~(\ref{e:Bernoul_gal}) does reduce to the
well-known Bernoulli theorem. Indeed, considering the coordinate system
$(x^\alpha)$ given by Eq.~(\ref{e:gab_weak}), the Killing vector $\vv{\xi}$
corresponds to the invariance by translation in the $t$ direction, so that we
have $\vv{\xi} = {\partial/\partial x^0} = c^{-1} {\partial/\partial t}$.
The components of $\vv{\xi}$ with respect to the coordinates $(x^\alpha)$
are thus simply
\be
	\xi^\alpha = (1,0,0,0) . 
\ee
Accordingly
\be \label{e:Bernoulli_Newt_prov}
	h\, \vv{\xi}\cdot\vv{u} = h \, u_\alpha \xi^\alpha = h \,  u_0
	\simeq - m_{\rm b} c^2 \left( 1 + \frac{H}{c^2} \right) 
		\left( 1 + \frac{\Phi}{c^2}  + \frac{v_j v^j}{2 c^2} \right) ,
\ee
where we have used Eq.~(\ref{e:h_Newt_lim}) for $h$ and Eq.~(\ref{e:ucova_Newt_lim})
for $u_0$. Expanding (\ref{e:Bernoulli_Newt_prov}) to first order in $c^{-2}$, we get
\be
	h\, \vv{\xi}\cdot\vv{u} \simeq - m_{\rm b} 
	\left( c^2 + H + \frac{1}{2} v_j v^j + \Phi \right) .
\ee
Since thanks to Eq.~(\ref{e:ua_Newt_lim}), 
\be
	\w{\nabla}_{\vv{u}} (h\, \vv{\xi}\cdot\vv{u}) = 
	u^\alpha \der{}{x^\alpha} (h\, \vv{\xi}\cdot\vv{u}) 
	= \frac{u^0}{c} \underbrace{\der{}{t}  (h\, \vv{\xi}\cdot\vv{u})}_{=0}
	+ \frac{v^i}{c} \der{}{x^i} (h\, \vv{\xi}\cdot\vv{u}) ,
\ee
we conclude that the Newtonian limit of Eq~(\ref{e:Bernoul_gal}) is
\be
	v^i \der{}{x^i} \left( H + \frac{1}{2} v_j v^j + \Phi \right) = 0 ,
\ee
i.e. we recover the classical Bernoulli theorem for a stationary flow.

\subsubsection{Axisymmetric flow}

In the case where the spacetime is axisymmetric (but not necessarily stationary),
there exists a coordinate system of spherical type $x^\alpha = (x^0=ct,r,\theta,\varphi)$
such that the Killing vector is
\be
	\vv{\xi} = \der{}{\varphi}
\ee
The conserved quantity $h\, \vv{\xi}\cdot\vv{u}$ is then interpretable as the 
\emph{angular momentum per baryon}.
Indeed its Newtonian limit is
\be
	h\, \vv{\xi}\cdot\vv{u} = h \, u_\alpha \xi^\alpha = 
	h \, u_\varphi \simeq m_{\rm b} c^2 \left( 1 + \frac{H}{c^2} \right) 
	\frac{v_\phi}{c} \simeq  m_{\rm b} c v_\varphi , 
\ee
where we have used Eq.~(\ref{e:ucova_Newt_lim}) to replace $u_\varphi$ by 
$v_\varphi/c$. In terms of the components $v_{(i)}$
of the fluid velocity in an orthonormal frame, one has 
$v_\varphi = r\sin\theta\, v_{(\varphi)}$, so that 
\be
	h\, \vv{\xi}\cdot\vv{u} = c \times r\sin\theta\, m_{\rm b} v_{(\varphi)} .
\ee
Hence, up to a factor $c$, the conserved quantity is the $z$-component of the 
angular momentum of one baryon.

\subsection{Irrotational flow}

A simple fluid is said to be \emph{irrotational} iff its vorticity 2-form vanishes
identically:
\be \label{e:w_zero_irrot}
	\encadre{ \w{w} = 0 } .
\ee
It is easy to see that this implies the vanishing of the 
\emph{kinematical vorticity vector} $\vv{\omega}$ defined by
\be \label{def_kin_vort}
	\omega^\alpha :=  \frac{1}{2} \epsilon^{\alpha\mu\rho\sigma}
	u_\mu \nabla_\rho u_\sigma 
	= \frac{1}{2} \epsilon^{\alpha\mu\rho\sigma}
	u_\mu \der{u_\sigma}{x^\rho} .
\ee
In this formula, $\epsilon^{\alpha\mu\rho\sigma}$ stands for 
the components of the alternating
type-$(4,0)$ tensor $\w{\bar\epsilon}$ that is related to the volume element 4-form
$\w{\epsilon}$  associated with $\w{g}$
by $\epsilon^{\alpha\beta\gamma\delta} \epsilon_{\alpha\beta\gamma\delta}=-4!$.
Equivalently, $\w{\bar\epsilon}$ is such that for any basis of 1-forms
$(\w{e}^\alpha)$ dual to a right-handed orthonormal vector basis 
$(\vv{e}_\alpha)$, then $\w{\bar\epsilon}(\w{e}^0,\w{e}^1,\w{e}^2,\w{e}^3)=1$.
Notice that the second equality in Eq.~(\ref{def_kin_vort}) results from 
the antisymmetry of $\w{\bar\epsilon}$ combined with the symmetry of the
Christoffel symbols in their lower indices. From the alternating character 
of $\w{\bar\epsilon}$, the kinematical vorticity vector $\vv{\omega}$ is by
construction orthogonal to the 4-velocity:
\be
	\vv{u} \cdot \vv{\omega} = 0 .
\ee
Moreover, at the non-relativistic limit, $\vv{\omega}$ is nothing but
the curl of the fluid velocity:
\be \label{e:curlv_Newt}
	\vv{\omega} \simeq \frac{1}{c} \mathrm{curl}\, \vec{v} .
\ee
That $\w{w}=0$ implies $\vv{\omega}=0$, as stated above, results from the relation
\be
	\omega^\alpha = \frac{1}{4h} \epsilon^{\alpha\mu\rho\sigma} u_\mu w_{\rho\sigma} ,
\ee
which is an easy consequence of 
$w_{\rho\sigma} = \partial(h\, u_\sigma)/\partial{x^\rho} -
 \partial(h\, u_\sigma)/\partial{x^\sigma}$ [Eq.~(\ref{e:def_vorticity})].
From a geometrical point of view, the vanishing of $\vv{\omega}$ implies that
the fluid worldlines are orthogonal to a family of (spacelike) hypersurfaces
(submanifolds of $\M$ of dimension 3).

The vanishing of the vorticity 2-form for an irrotational fluid, 
Eq.~(\ref{e:w_zero_irrot}), means that 
that the fluid momentum per baryon 1-form $\w{\pi}$ is closed: 
$\dd\w{\pi}=0$. By Poincar\'e lemma (cf. Sec.~\ref{s:diff_forms}), there exists
then a scalar field $\Psi$ such that
\be \label{e:h_u_dpsi}
	\encadre{ \w{\pi} = \dd \Psi }, \qquad 
	\mbox{i.e.} \qquad \encadre{ h\, \uu{u} = \dd \Psi } .
\ee
The scalar field $\Psi$ is called the \emph{potential} of the flow.
Notice the difference with the Newtonian case: a relativistic irrotational flow is
such that $h\, \uu{u}$ is a gradient, not $\uu{u}$ alone. Of course
at the Newtonian limit $h\rightarrow m_{\rm b}c^2 = {\rm const}$, so that the two
properties coincide.

For an irrotational fluid, the Carter-Lichnerowicz equation of motion 
(\ref{e:CL_sf}) reduces to
\be
	T \dd {\bar s}  = 0 .
\ee
Hence the fluid must either have a zero temperature or be isentropic.
The constraint on $\Psi$ arises from the baryon number conservation,
Eq.~(\ref{e:baryon_cons}). Indeed, we deduce from Eq.~(\ref{e:h_u_dpsi}) that
\be \label{e:u_dPsi}
	\vv{u} = \frac{1}{h} \vv{\nabla} \Psi,
\ee
where $\vv{\nabla} \Psi$ denotes the vector associated to the gradient 1-form
$\dd\Psi = \w{\nabla}\Psi$ by the standard metric duality, the components
of $\vv{\nabla} \Psi$ being $\nabla^\alpha\Psi = g^{\alpha\mu} \nabla_\mu \Psi$.
Inserting (\ref{e:u_dPsi}) into the baryon number conservation equation (\ref{e:baryon_cons}) yields
\be \label{e:dal_Psi}
	\frac{n}{h} \square \Psi + \w{\nabla} \left(\frac{n}{h}\right) \cdot
	\vv{\nabla} \Psi = 0 ,
\ee
where $\square$ is the d'Alembertian operator associated with the metric $\w{g}$:
$\square := \w{\nabla}\cdot\vv{\nabla} = \nabla_\mu \nabla^\mu = g^{\mu\nu} \nabla_\mu
\nabla_\nu$.

Let us now suppose that the spacetime possesses some symmetry described by the Killing
vector $\vv{\xi}$. Then Eq.~(\ref{e:xi_w}) applies. Since $\w{w}=0$ in the
present case, it reduces to 
\be
   \dd \langle\w{\pi},\vv{\xi}\rangle = 0 .
\ee
We conclude that the scalar field
$\langle\w{\pi},\vv{\xi}\rangle$ is constant, or equivalently
\be\label{e:int_prem_irrot}
	\encadre{ h\, \vv{\xi}\cdot\vv{u} = {\rm const.} }
\ee
Hence for an irrotational flow, the quantity $h\, \vv{\xi}\cdot\vv{u}$
is a global constant, and not merely a constant along each fluid line which may vary
from a fluid line to another one.
One says that $h\, \vv{\xi}\cdot\vv{u}$ is a \emph{first integral of motion}.
This property of irrotational relativistic fluids was first established 
by Lichnerowicz (\cite{Lichn41}), the special relativistic subcase (flat spacetime)
having been proved previously by Synge (\cite{Synge37}).

\begin{figure}
\centerline{\includegraphics[height=0.38\textheight]{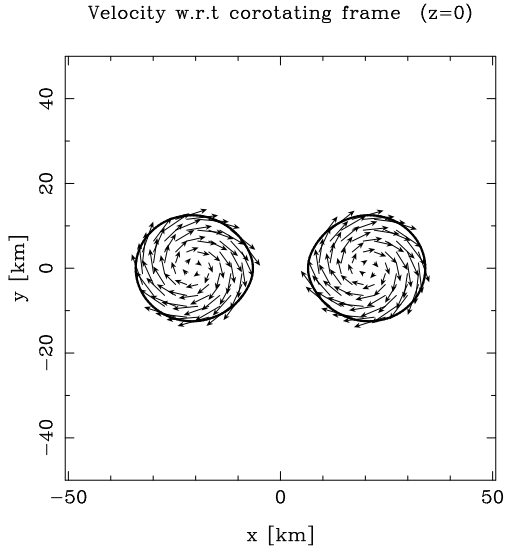}}
\caption{\label{f:irrot_bin_ns} Velocity with respect to a co-orbiting observer 
for irrotational binary relativistic stars. The figure is drawn in the orbital plane. 
The velocity field  has been obtained
by numerically solving Eq.~(\ref{e:dal_Psi}) for the fluid potential $\Psi$.
The first integral of motion (\ref{e:int_prem_irrot}) provided by the
helical Killing vector has been used to get the enthalpy per baryon $h$. The density profile in the stars is then deduced from the EOS
[from (Gourgoulhon \etal\ \cite{GourgGTMB01})].}
\end{figure}

As an illustration of the use of the integral of motion (\ref{e:int_prem_irrot}), 
let us consider the problem of equilibrium configurations of irrotational relativistic stars in binary systems. This problem is particularly relevant for describing the last stages of the slow inspiral of binary neutron stars, which are expected to be one of the strongest sources of gravitational waves for the interferometric detectors LIGO, GEO600, and VIRGO (see Baumgarte \& Shapiro (\cite{BaumgS03}) for a review about relativistic
binary systems).
Indeed the shear viscosity of nuclear matter is not sufficient to synchronize the
spin of each star with the orbital motion within the short timescale of the 
gravitational radiation-driven inspiral. Therefore contrary to ordinary stars, close
binary system of neutron stars are not in synchronized rotation. Rather if the spin
frequency of each neutron star is initially low (typically $1 {\rm\  Hz}$), the
orbital frequency in the last stages is so high (in the kHz regime), that it is a
good approximation to consider that the fluid in each star is irrotational.
Besides, the spacetime containing an orbiting binary system has a priori no
symmetry, due to the emission of gravitational wave. However, in the inspiral phase,
one may approximate the evolution of the system by a sequence of equilibrium
configurations consisting of exactly circular orbits. Such a configuration possesses
a Killing vector, which is helical, being of the type
$\vv{\xi} = \partial/\partial t + \Omega\partial/\partial\varphi$, where $\Omega$
is the orbital angular velocity (see Friedman \etal\ \cite{FriedUS02} for more details).
This Killing vector provides the first integral
of motion (\ref{e:int_prem_irrot}) which permits to solve the problem 
(cf. Fig.~\ref{f:irrot_bin_ns}). It is worth noticing that the derivation
of the first integral of motion directly from the relativistic Euler equation
(\ref{e:Euler_relat_u}), i.e. without using the Carter-Lichnerowicz equation,
is quite lengthy (Shibata \cite{Shiba98}, Teukolsky \cite{Teuko98}).

\subsection{Rigid motion}

Another interesting case in which there exists a first integral of motion
is that of a rigid isentropic flow. We say that the fluid is in \emph{rigid motion} iff
(i) there exists a Killing vector $\vv{\xi}$ and (ii) the fluid 4-velocity is collinear
to that vector:
\be \label{e:u_rigid}
	\encadre{ \vv{u} = \lambda \, \vv{\xi} } ,
\ee
where $\lambda$ is some scalar field (not assumed to be constant).
Notice that this relation implies that the Killing vector $\vv{\xi}$ is timelike
in the region occupied by the fluid. Moreover, the normalization relation
$\vv{u}\cdot\vv{u}=-1$ implies that $\lambda$ is related to the scalar square
of $\vv{\xi}$ via
\be
	\lambda = \left( -\vv{\xi}\cdot\vv{\xi} \right) ^{1/2} .
\ee
The denomination \emph{rigid} stems from the fact that (\ref{e:u_rigid}) in conjunction
with the Killing equation (\ref{e:Killing_eq}) implies\footnote{the reverse has been proved to hold for an isentropic fluid (Salzman \& Taub \cite{SalzmT54})} that both the expansion rate
$\theta:=\w{\nabla}\cdot\vv{u}$ [cf. Eq.~(\ref{e:expansion_rate})] and the shear tensor
(see e.g. Ehlers \cite{Ehler61})
\be
	\sigma_{\alpha\beta}:= \frac{1}{2}
		(\nabla_\mu u_\nu + \nabla_\nu u_\mu) P^\mu_{\ \, \alpha} P^\nu_{\ \, \beta}
	- \frac{1}{3}  \theta P_{\alpha\beta}
\ee
vanish identically for such a fluid.

Equation~(\ref{e:xi_w}) along with Eq.~(\ref{e:u_rigid}) results in
\be 
	\frac{1}{\lambda}\vv{u}\cdot\w{w} + \dd \langle\w{\pi},\vv{\xi}\rangle = 0 .
\ee
Then, the Carter-Lichnerowicz equation (\ref{e:CL_sf}) yields
\be 
	\frac{T}{\lambda}\dd{\bar s} + \dd\langle\w{\pi},\vv{\xi}\rangle = 0 .
\ee
If we assume that the fluid is isentropic, $\dd{\bar s}=0$ and we get the same first integral of motion than in the irrotational case:
\be
	\encadre{ h\, \vv{\xi}\cdot\vv{u} = {\rm const.} }
\ee
 
This first integral of motion has been massively used to compute 
stationary and axisymmetric configurations of rotating stars in general
relativity (see Stergioulas \cite{Sterg03} for a review). In this case, the
Killing vector $\vv{\xi}$ is 
\be
	\vv{\xi} = \vv{\xi}_{\rm station} + \Omega\, \vv{\xi}_{\rm axisym}, 
\ee
where $\vv{\xi}_{\rm station}$ and $\vv{\xi}_{\rm axisym}$ are the Killing vectors
associated with respectively stationarity and axisymmetry. Note that the isentropic
assumption is excellent for neutron stars which are cold objects.

\subsection{First integral of motion in symmetric spacetimes}

The irrotational motion in presence of a Killing vector $\vv{\xi}$ and
the isentropic rigid motion treated above are actually subcases of flows that satisfy
the condition 
\be \label{e:xi_w_first_int}
	\encadre{ \vv{\xi}\cdot\w{w} = 0 } ,
\ee
which is necessary and sufficient for 
$\langle\w{\pi},\vv{\xi}\rangle = h\, \vv{\xi}\cdot\vv{u}$ to be a first integral of motion.
This property follows immediately from Eq.~(\ref{e:xi_w}) (i.e. 
the symmetry property $\Lie{\xi}\w{\pi}=0$ re-expressed  via the Cartan identity)
and was first noticed by Lichnerowicz (\cite{Lichn55}).
For an irrotational motion, Eq.~(\ref{e:xi_w_first_int}) holds trivially because
$\w{w}=0$, whereas for an isentropic rigid motion it holds thanks to the
isentropic Carter-Lichnerowicz equation of motion (\ref{e:CL_baro}) with
$\vv{u} = \lambda \vv{\xi}$.

\subsection{Relativistic Kelvin theorem}

Here we do no longer suppose that the spacetime has any symmetry.
The only restriction that we set is  that
the fluid must be isentropic, as discussed in Sec.~\ref{s:isentropic}.
The Carter-Lichnerowicz equation of motion (\ref{e:CL_baro}) leads then very
easily to a relativistic generalization of Kelvin theorem about conservation
of circulation.
Indeed, if we apply Cartan identity (\ref{e:Cartan}) to express the Lie derivative
of the fluid vorticity 2-form $\w{w}$
along the vector $\alpha\vv{u}$ (where $\alpha$ is any non-vanishing scalar field), 
we get
\be
	\w{\mathcal{L}}_{\alpha\vv{u}}\, \w{w} = 
	\alpha\vv{u}\cdot \underbrace{\dd\w{w}}_{=0}
	+ \dd( \alpha\underbrace{\vv{u}\cdot\w{w}}_{=0}) .
\ee
The first ``$=0$'' results from $\dd\w{w} = \dd\dd\w{\pi}=0$ [nilpotent character
of the exterior derivative, cf. Eq.~(\ref{e:ext_der_nilpot})], whereas the 
second ``$=0$'' is the isentropic Carter-Lichnerowicz equation (\ref{e:CL_baro}).
Hence
\be \label{e:lieu_w_zero}
	\encadre{ \w{\mathcal{L}}_{\alpha\vv{u}}\, \w{w} = 0 } .
\ee
This constitutes a relativistic generalization of Helmholtz's vorticity equation
\be
	\der{\vec{\omega}}{t} = \vec{\omega}\cdot\vec{\nabla}\vec{v}
	- \vec{v}\cdot\vec{\nabla}\vec{\omega} - (\mathrm{div}\, \vec{v})\, \vec{\omega} ,
\ee
which governs the evolution of $\vec{\omega}:=\mathrm{curl}\, \vec{v}$ 
[cf. Eq.~(\ref{e:curlv_Newt})].

The \emph{fluid circulation} around a closed curve $C\subset\M$ is defined
as the integral of the fluid momentum per baryon along $C$:
\be \label{e:def_circulation}
	\encadre{ \mathcal{C}(C) := \int_C \w{\pi} } .
\ee
Let us recall that $C$ being a 1-dimensional manifold and $\w{\pi}$ a 1-form the
above integral is well defined, independently of any length element on $C$.
However to make the link with traditional notations in classical hydrodynamics,
we may write $\w{\pi} = h\, \uu{u}$ [Eq.~(\ref{e:pi_hu})] and let appear the vector
$\vv{u}$ (4-velocity) associated to the 1-form $\uu{u}$ by the standard
metric duality. Hence we can rewrite (\ref{e:def_circulation}) as
\be \label{e:circulation_alt}
	\mathcal{C}(C) = \int_C h\, \vv{u}\cdot d\vv{\ell} .
\ee
This writing makes an explicit use of the metric tensor $\w{g}$ (in the scalar
product between $\vv{u}$ and the small displacement $d\vv{\ell}$).

Let $S$ be a (2-dimensional) compact surface the boundary of which is $C$:
$C = \partial S$. Then by the Stokes theorem (\ref{e:Stokes}) and the 
definition of $\w{w}$,
\be
	\mathcal{C}(C) = \int_S \dd\w{\pi} = \int_S \w{w} .
\ee

We consider now that the loop $C$ is dragged along the fluid worldlines. 
This means that we consider a 1-parameter family of loops $C(\lambda)$
that is generated from a initial loop $C(0)$ nowhere tangent to $\vv{u}$
by displacing each point of $C(0)$ by some distance along the field lines
of $\vv{u}$. We consider as well a family of surfaces $S(\lambda)$ such that
$\partial S(\lambda) = C(\lambda)$. We can parametrize each fluid worldline 
that is cut by $S(\lambda)$ by the parameter $\lambda$ instead of the proper time $\tau$.
The corresponding tangent vector is then $\vv{v} = \alpha\vv{u}$, where
$\alpha := d\tau/d\lambda$ (the derivative being taken along a given fluid worldline).
From the very definition of the Lie derivative (cf. Sec.~\ref{s:lie_deriv} where the Lie derivative of a vector has been
defined from the dragging of the vector along the flow lines),
\be
	\frac{d}{d\lambda} \mathcal{C}(C) = \frac{d}{d\lambda} \int_S \w{w}
	= \int_S  \w{\mathcal{L}}_{\alpha\vv{u}}\, \w{w} .
\ee
From Eq.~(\ref{e:lieu_w_zero}), we conclude
\be \label{e:Kelvin}
	\encadre{ \frac{d}{d\lambda} \mathcal{C}(C) = 0 } .
\ee
This is the \emph{relativistic Kelvin theorem}.
It is very easy to show that in the Newtonian limit it reduces to the classical
Kelvin theorem. Indeed, choosing $\lambda=\tau$, the  
non-relativistic limit yields $\tau=t$, where $t$ is the absolute
time of Newtonian physics. Then each curve $C(t)$ lies in the hypersurface $t={\rm const}$
(the ``space'' at the instant $t$, cf. Sec.~\ref{s:curved_spacetime}). Consequently, the scalar product $\vv{u}\cdot d\vv{\ell}$ in (\ref{e:circulation_alt}) involves only
the spatial components of $\vv{u}$, which according to Eq.~(\ref{e:ua_Newt_lim})
are $u^i \simeq v^i/c$. Moreover the Newtonian limit of $h$ is
$m_{\rm b} c^2$ [cf. Eq.~(\ref{e:h_Newt_lim})], so that (\ref{e:circulation_alt})
becomes
\be
	\mathcal{C}(C) \simeq m_{\rm b} c \int_C \vec{v}\cdot d\vec{\ell} .
\ee
Up to the constant factor $m_{\rm b} c$ we recognize the classical expression
for the fluid circulation around the circuit $C$.
Equation (\ref{e:Kelvin}) reduces than to the classical Kelvin theorem
expressing the constancy of the fluid circulation around a closed loop which 
is comoving with the fluid.

\subsection{Other conservation laws}

The Carter-Lichnerowicz equation enables one to get easily other relativistic
conservation laws, such as the conservation of \emph{helicity} or 
the conservation of \emph{enstrophy}. We shall not discuss them in this
introductory lecture and
refer the reader to articles by Carter (\cite{Carte79}, \cite{Carte89}),
Katz (\cite{Katz84})
or Bekenstein (\cite{Beken87}).


\section{Conclusions}

The Carter-Lichnerowicz formulation is well adapted to 
a first course in relativistic hydrodynamics. Among other things,
it uses a clear separation between what is a vector and 
what is a 1-form, which has a deep physical significance
(as could also be seen from the variational formulations of
hydrodynamics mentioned in Sec.~\ref{s:divT_zero}). 
For instance, velocities are fundamentally vectors, whereas momenta
are fundamentally 1-forms.
On the contrary, the ``standard'' tensor calculus mixes very often
the concepts of vector and 1-form, via an immoderate use of the 
metric tensor. Moreover, we hope that the reader is now convinced that 
the Carter-Lichnerowicz approach greatly facilitates the derivation 
of conservation laws.
It must also be said that, although we have not discussed
it here, this formulation can be applied directly
to non-relativistic hydrodynamics, by introducing exterior calculus on the
Newtonian spacetime, and turns out to be very fruitful
(Carter \& Gaffet \cite{CarteG88}; Prix \cite{Prix04}; Carter \& Chamel \cite{CarteC04};
Chamel \cite{Chame04}).
Besides, it is worth to mention that the Carter-Lichnerowicz approach can also be extended
to relativistic magnetohydrodynamics (Lichnerowicz \cite{Lichn67} and Sec.~9 of
Carter \etal\ \cite{CarteCC06}).

In this introductory lecture, we have omitted important topics,
among which relativistic shock waves (see e.g. Mart\'\i\  \& M\"uller \cite{MartiM03};
Font \cite{Font03}, Anile \cite{Anil89}),
instabilities in rotating relativistic fluids (see e.g. Stergioulas \cite{Sterg03}; Andersson \cite{Ander03}; Villain~\cite{Villa06}),
and superfluidity (see e.g. Carter \& Langlois \cite{CarteL98},
Prix \etal\ \cite{PrixNC05}).
Also we have not discussed much astrophysical applications.
We may refer the interested reader to the review article by Font
(\cite{Font03}) for relativistic hydrodynamics in strong gravitational fields, to
Shibata \etal\ (\cite{ShibaTU05}) for some recent application to the merger
of binary neutron stars, and
to Baiotti \etal\ (\cite{Baiot_al05}), Dimmelmeier \etal\ (\cite{DimmeNFIM05}),
and Shibata \& Sekiguchi (\cite{ShibaS05}) for applications to gravitational collapse.
Regarding the treatment of relativistic jets, which requires only special relativity, 
we may mention Sauty \etal\ (\cite{SautyMTT04}) and
Mart\'\i\ \& M\"uller (\cite{MartiM03}) for reviews of respectively analytical 
and numerical approaches, as well as 
Alloy \& Rezzola (\cite{AlloyR06}) for an example of recent work.

\acknowledgements It is a pleasure to thank B\'erang\`ere
Dubrulle and Michel Rieutord for having organized the very successful Carg\`ese school
on Astrophysical Fluid Dynamics. 
I warmly thank Brandon Carter for fruitful discussions and for reading the manuscript.
I am extremely grateful to Silvano Bonazzola for having introduced
me to relativistic hydrodynamics (among other topics !), and for his constant stimulation and inspiration. In the spirit of the Carg\`ese school, I dedicate this article to him, recalling
that his very first scientific paper (Bonazzola \cite{Bonaz62})
regarded the link between physical measurements and geometrical operations 
in spacetime, like the orthogonal decomposition of the 4-velocity which we discussed in Sec.~\ref{s:worldlines}.


\begin{thebibliography}{99}

\bibitem[2006]{AlloyR06}
Alloy, M.A. \& Rezzolla, L. 2006, ApJ, in press
[preprint: astro-ph/0602437]

\bibitem[2003]{Ander03}
Andersson, N. 2003, Class. Quantum Grav. 20, R105

\bibitem[1989]{Anil89}
Anile, A.M. 1989, \emph{``Relativistic Fluids and Magneto-fluids''}
(Cambridge University Press, Cambridge)

\bibitem[2005]{Baiot_al05}
Baiotti, L., Hawke, I., Montero, P.J., L\"offler, F., Rezzolla, L., Stergioulas, N.,
Font, J.A., \& Seidel, E. 2005,
Phys. Rev. D 71, 024035

\bibitem[2003]{BaumgS03}
Baumgarte, T.W. \& Shapiro, S.L. 2003, Phys. Rep. 376, 41

\bibitem[1987]{Beken87}
Bekenstein, J.D 1987, ApJ 319, 207

\bibitem[1962]{Bonaz62}
Bonazzola, S. 1962, Nuovo Cimen. 26, 485
 
\bibitem[2004]{Carro04}
Carroll, S.M. 2004, \emph{``Spacetime and Geometry: An Introduction to General Relativity''}
(Addison Wesley / Pearson Education, San Fransisco)

\bibitem[1973]{Carte73}
Carter, B. 1973, Commun. Math. Phys. 30, 261 

\bibitem[1979]{Carte79}
Carter, B. 1979,
in \emph{``Active Galactic Nuclei''},
ed. C.~Hazard \& S.~Mitton (Cambridge University Press, Cambridge), p.~273

\bibitem[1989]{Carte89}
Carter, B. 1989,
in \emph{``Relativistic Fluid Dynamics''}, 
ed. A.~Anile \& Y. Choquet-Bruhat, 
Lecture Notes In Mathematics 1385 (Springer, Berlin), p.~1

\bibitem[2006]{CarteCC06}
Carter, B., Chachoua, E., \& Chamel, N. 2006,
Gen. Relat. Grav. 38, 83

\bibitem[2004]{CarteC04}
Carter, B. \& Chamel, N. 2004,
Int. J. Mod. Phys. D 13, 291

\bibitem[1988]{CarteG88}
Carter, B. \& Gaffet, B. 1988, J. Fluid. Mech. 186, 1

\bibitem[1998]{CarteL98}
Carter, B. \& Langlois, D. 1998, Nucl. Phys. B 531, 478

\bibitem[2004]{Chame04}
Chamel, N. 2004 \emph{``Entra\^\i nement dans l'\'ecorce d'une \'etoile \`a neutrons''},
PhD thesis, Universit\'e Paris 6

\bibitem[1993]{ComerL93}
Comer, G.L. \& Langlois, D. 1993, Class. Quantum Grav. 10, 2317

\bibitem[2005]{DimmeNFIM05}
Dimmelmeier, H., Novak, J., Font, J.A., Ib\'a\~nez, J.M., \& M\"uller, E. 2005,
Phys. Rev. D 71, 064023 

\bibitem[1961]{Ehler61}
Ehlers, J. 1961, Abhandl. Akad. Wiss. Mainz. Math. Naturw. Kl. 11, 792
[English translation in Gen. Relat. Grav. 25, 1225 (1993)]

\bibitem[2003]{Font03}
Font, J.A. 2003, Living Rev. Relativity 6, 4, 
\texttt{http://www.livingreviews.org/lrr-2003-4}

\bibitem[2002]{FriedUS02}
Friedman, J.L., Uryu, K., \& Shibata, M. 2002,
Phys. Rev. D 65, 064035

\bibitem[2001]{GourgGTMB01}
Gourgoulhon, E., Grandcl\'ement, P., Taniguchi, K., Marck, J.-A., \&
Bonazzola, S. 2001, 
Phys. Rev. D 63, 064029

\bibitem[2003]{Hartl03}
Hartle, J.B. 2003, \emph{``Gravity: An Introduction to Einstein's General Relativity''} (Addison Wesley / Pearson Education, San Fransisco)

\bibitem[1984]{Katz84}
Katz, J. 1984, Proc. R. Soc. Lond. A 391, 415

\bibitem[1940]{Lichn40}
Lichnerowicz, A. 1940, C. R. Acad. Sci. Paris 211, 117

\bibitem[1941]{Lichn41}
Lichnerowicz, A. 1941, Ann. Sci. \'Ecole Norm. Sup. 58, 285
[freely available from \texttt{http://www.numdam.org/}]

\bibitem[1955]{Lichn55}
Lichnerowicz, A. 1955, \emph{``Th\'eories relativistes de la gravitation et de
l'\'electromagn\'etisme''} (Masson, Paris)

\bibitem[1967]{Lichn67}
Lichnerowicz, A. 1967,  \emph{``Relativistic hydrodynamics and magnetohydrodynamics''}
(Benjamin, New York)

\bibitem[2003]{MartiM03}
Mart\'\i, J.M. \& M\"uller, E. 2003,
Living Rev. Relativity 6, 7, \\
\texttt{http://www.livingreviews.org/lrr-2003-7}

\bibitem[2004]{Prix04}
Prix, R. 2004, Phys. Rev. D 69, 043001

\bibitem[2005]{PrixNC05}
Prix, R., Novak, J., \&  Comer, G.L. 2005, 
Phys. Rev. D 71, 043005

\bibitem[1997]{Rieut97}
Rieutord, M. 1997, \emph{``Une introduction \`a la dynamique des fluides''}
(Masson, Paris)

\bibitem[1954]{SalzmT54}
Salzman, G. \& Taub, A.H. 1954, Phys. Rev. 95, 1659

\bibitem[2004]{SautyMTT04}
Sauty, C., Meliani, Z., Trussoni, E., \& Tsinganos, K. 2004, 
in \emph{``Virtual astrophysical jets''}, ed. S. Massaglia, G. Bodo \&
P. Rossi (Kluwer Academic Publishers, Dordrecht), p.~75

\bibitem[1998]{Shiba98}
Shibata, M. 1998, Phys. Rev. D 58, 024012

\bibitem[2005]{ShibaS05}
Shibata, M. \& Sekiguchi, Y. 2005,
Phys. Rev. D 71, 024014

\bibitem[2005]{ShibaTU05}
Shibata, M., Taniguchi, K., \& Uryu, K. 2005,
Phys. Rev. D 71, 084021

\bibitem[2003]{Sterg03}
Stergioulas, N. 2003, Living Rev. Relativity 6, 3, \\
\texttt{http://www.livingreviews.org/lrr-2003-3}

\bibitem[1937]{Synge37}
Synge, J.L 1937, Proc. London Math. Soc. 43, 376
[reprinted in Gen. Relat. Grav. 34, 2177 (2002)]

\bibitem[1954]{Taub54}
Taub, A.H. 1954, Phys. Rev. 94, 1468

\bibitem[1998]{Teuko98}
Teukolsky, S.A 1998, ApJ 504, 442

\bibitem[2006]{Villa06}
Villain, L. 2006, this volume (also preprint astro-ph/0602234)

\bibitem[1984]{Wald84}
Wald, R.M. 1984, \emph{``General Relativity}, 
(Univ. Chicago Press, Chicago)

\end{thebibliography}
\end{document}